# A nanopore-gated sub-attoliter silicon nanocavity for single-molecule trapping and analysis

**Funing Liu[1], Qitao Hu[1,6], Anton Sabantsev[2], Giovanni Di Muccio[3,4], Shuangshuang Zeng[1,7], Mauro Chinappi[5], Sebastian Deindl[2] and Zhen Zhang[1]**

[1]Division of Solid-State Electronics, Department of Electrical Engineering, Uppsala University, BOX 65, SE-75121, Uppsala, Sweden.
[2]Department of Cell and Molecular Biology, Science for Life Laboratory, Uppsala University, Uppsala, Sweden.
[3]NY-Masbic, Department of Life and Environmental Sciences, Università Politecnica delle Marche, Via Brecce Bianche, 60131 Ancona, Italy
[4]National Future Biodiversity Centre (NFBC), Palermo, Italy
[5]Department of Industrial Engineering, University of Rome Tor Vergata, Roma, Italy.
[6]Current address: Department of Radiology, Stanford University, Stanford, CA 94305 USA.
[7]Current address: School of Integrated Circuits, Huazhong University of Science and Technology, Wuhan 430074, China.

Correspondence should be addressed to S.D. (sebastian.deindl@icm.uu.se) and Z.Z. (zhen.zhang@angstrom.uu.se).

**Biomolecules exhibit dynamic conformations critical to their functions, yet observing these processes at the single-molecule level under native conditions remains a formidable challenge. While surface immobilization has been widely used to extend observation times, it could disrupt molecular dynamics and impede biological function. Moreover, the study of weak molecular interactions requires high local concentrations, often leading to problems with signal saturation in fluorescence-based approaches. Recent advancements in single-molecule trapping techniques have addressed some limitations, but achieving precise, controllable, long-term trapping in a molecularly crowded environment without external forces remains difficult. Here, we introduce a nanopore-gated sub-attoliter silicon nanocavity that enables precise, non-perturbative trapping of individual biomolecules for extended observation times, eliminating the need for external forces. Using nucleosomes as model systems, we demonstrate single-molecule Förster resonance energy transfer (smFRET) to monitor relative distances. With smFRET, we directly observe dynamic unwrapping and rewrapping events induced by the chromatin remodeling enzyme Chd1, as well as weak interactions between two nucleosomes trapped inside the nanocavity. To further demonstrate the versatility of our device for studying weak molecular interactions, we directly observed, at the single-molecule level, interactions of the prototypical transcription factor LacI with a weak operator under near-physiological salt conditions. Our data also show that an applied electric field can modulate the conformational properties of macromolecules, emphasizing a key advantage of our device: it does not require an electric field to retain trapped molecules. We envision this nanocavity platform as a powerful tool for interrogating molecular dynamics in physiologically relevant environments, offering unperturbed access to weak and transient interactions that are central to biological regulation.**

# Main

Biomolecules adopt diverse conformations essential for their functions. Structural biology reveals these states through high-resolution techniques like cryo-EM, which captures distinct conformations in macromolecular ensembles. While static snapshots reveal equilibrium states, they fail to capture the dynamics of state interconversion. In ensemble measurements, these transitions are often obscured due to asynchronous molecular behaviour. Single-molecule techniques have revolutionized the study of complex biomolecular processes, enabling the observation of dynamic conformational changes and transient interactions that are often masked in ensemble measurements[1–19]. A central challenge remains the need to observe single molecules over extended periods under conditions that preserve their native behaviour[20–25].

A widely used approach for extending observation times is to anchor biomolecules to a surface, such as a coverslip[26]. Similarly, optical tweezers can trap single molecules by holding a bead, conjugated to the molecule, in place with a highly focused laser beam[27]. While surface-immobilization-based methods have unquestionably yielded highly significant results and continue to evolve[28,29], the immobilization process could disrupt the molecule's natural dynamics and, in some cases, impede its biological function[30–32]. Even when such perturbations are minimal, these methods remain limited in their ability to study weak molecular interactions that demand high local concentrations, primarily due to signal-saturation issue in fluorescence-based approaches[33]. To overcome these limitations, several techniques have been developed in recent years for the longer-term observation of single molecules without immobilization. Methods such as the nanopore electro-osmotic trap (NEOtrap)[34] utilize electro-osmotic forces to trap single molecules exploiting a DNA-origami nanosphere docked on a nanopore, allowing observation for hours. Similarly, the electrokinetic nanovalve relies on dielectrophoretic forces in lab-on-chip setups to guide and confine molecules[35], while the anti-Brownian electrophoretic trap (ABEL trap) uses real-time feedback voltage to counteract Brownian motion, enabling trapping in open environments[23,24,36–39]. However, these elegant approaches depend on external electric fields, which could potentially perturb the molecule's native dynamics[40–42]. Electrostatic fluid traps[43–45] and porous vesicle encapsulation techniques[46,47] allow passive trapping of single molecules without the application of external forces. Despite their simplicity, these methods rely on stochastic loading, making precise control difficult and limiting their applicability to systems requiring selective or repeatable trapping. Entropic cages offer a promising, alternative, confining single molecules through spatial constraints without external forces[48,49]. Although this approach is gentle and highly controllable, it is currently limited to large biomolecules, such as very long DNA molecules, due to the size of the apertures and cages. This limits its use for smaller biomolecules and complexes.

Despite these significant advancements, achieving a solution that fulfils all desired capabilities - controllable, non-perturbative trapping of single molecules with extended observation times in a molecularly crowded environment[50,51] (similar to the intracellular space, which often features high molecular concentrations) - remains a challenge. Here, we present a novel nanopore-gated sub-attoliter silicon nanocavity device as a robust entropic trapping platform for non-perturbative single-molecule trapping and analysis. Fabricated in a thin silicon membrane of a silicon-on-insulator (SOI) wafer using high-precision silicon processing, the nanocavity connects two liquid reservoirs through the *trans/cis* nanopore gates of different sizes, enabling ionic current flow under an applied electrical bias. The larger nanopore allows controlled electrical loading or release of single molecules between the nanocavity and one reservoir, while the smaller nanopore prevents unintended molecule escape. Real-time monitoring of ionic current can provide feedback for precise molecule loading. Once loaded, the electrical bias is switched off, and the molecule is entropically trapped for analysis. To fine-tune nanopore sizes for specific molecules, we developed a customized carbon deposition technique. Using nanopores of 12 nm and 9 nm, we demonstrated the controlled loading, release, and trapping of single or multiple 11 nm-sized nucleosomes. Furthermore, single-molecule Förster resonance energy transfer (smFRET)[52–54] analysis within the nanodevice reveals the dynamics of trapped nucleosomes and highlights the modulatory effects of applied electric fields. We also used the device to directly observe stacking interactions between two nucleosomes co-trapped in the nanocavity, demonstrating its ability to trap biomolecules precisely and non-perturbatively for extended single-molecule observation in a molecularly crowded environment. To further demonstrate the advantage of nanoscale confinement for studying weak molecular interactions, we directly observed, at the single-molecule level, interactions of the prototypical transcription factor LacI with a weak operator under near-physiological salt conditions.

**Nanocavity device design and operation**

To enable controllable trapping of individual molecules in a non-perturbative environment, we designed a nanocavity device comprising a top silicon nanopore (*trans*), a sub-attoliter silicon nanocavity, and a bottom silicon nitride ($SiN_x$) nanopore (*cis*). We fabricated the device in a free-standing silicon membrane of a silicon-on-insulator (SOI) wafer, with membrane and $SiN_x$ layers 88 nm and 30 nm thick, respectively (Fig. 1a, Supplementary Fig. 1). To reduce background signal, we deposited a 50-nm-thick gold layer onto the $SiN_x$ layer, 1 μm beyond the $SiN_x$ nanopore. Adjusting membrane thickness and wet-etching duration[55] allowed precise control over the nanocavity size. Moreover, we developed a carbon deposition technique to precisely reduce nanopore sizes. To characterize the resulting nanocavity device, we used cross-sectional scanning electron microscopy (SEM) before carbon deposition and transmission electron microscopy (TEM) after carbon deposition, revealing a 12-nm silicon nanopore and a 9-nm $SiN_x$ nanopore (Fig. 1b). These dimensions were specifically chosen to accommodate the target molecule investigated in this work. In addition, the solid surfaces of the nanocavity were passivated with BSA prior to the loading of actual target molecules to prevent surface sticking.

The thin membrane separates the imaging buffer containing electrolyte solution into *trans* and *cis* reservoirs (Fig. 1a, b). The only fluid pathway connecting these reservoirs passes through the silicon nanopore, the nanocavity, and the $SiN_x$ nanopore. Accordingly, applying a voltage between the reservoirs *via* Ag/AgCl electrodes is expected to generate an ionic current through this pathway. We reason that an electrical bias could pull molecules from the *trans* reservoir through the silicon nanopore into the nanocavity *via* electrophoretic[56] or electroosmotic[57] forces. A fluorescence microscope beneath the *cis* reservoir allows visualization of fluorophore-labelled molecules in the nanocavity (Fig. 1a).

To assess the performance of our device using a biologically relevant macromolecular complex, we reconstituted mononucleosomes comprising histone octamers labelled with the FRET donor dye Cy3 on histone H2A(K120C) and double-stranded DNA labelled with the acceptor dye Cy5 at the short linker end (Fig. 1c). The nucleosome, the fundamental packaging unit of eukaryotic genomes, consists of a histone octamer wrapped by approximately 147 base pairs (bp) of DNA[58]

Its positioning, modifications, and dynamics are critical for regulating genome accessibility and function. Furthermore, we used our device to create an effective local concentration sufficiently high to enable the observation of interactions between the iconic transcription factor LacI and a weak operator at near-physiological salt concentrations without saturating the fluorescence readout.

**Customizable nanodevices with tunable nanopore gate sizes**

To determine the optimal nanopore gate sizes for trapping FRET-labelled nucleosomes, we considered the following: if both the *trans* and *cis* nanopores were larger than the target molecule, the molecule would translocate through the nanocavity under electrical bias. If the *trans* nanopore were much larger and the *cis* nanopore smaller than the target molecule, the molecule would only be held within the nanocavity under electrical bias and could, upon bias removal, easily escape *via* the *trans* nanopore. Conversely, if the *trans* nanopore could match the target molecule size and the *cis* nanopore were smaller, the molecule would be loaded into the nanocavity by electrical force and remain entropically trapped without an external field. Initial tests on the relationship between nanopore gate size and the dwell time of 20 kb DNA confirmed that appropriately sized nanopores are critical for prolonged trapping (Devices a-c in Supplementary Fig. 2). Given the nucleosome diameter of ~11 nm[58], we further refined nanopore dimensions using electron beam-induced carbon deposition (Supplementary Fig. 3). This method enabled the fabrication of several devices with defined I–V characteristics (Supplementary Fig. 4), including one with *trans* and *cis* nanopores of 12 nm and 9 nm (Device 1), respectively—dimensions ideally suited for trapping nucleosomes (Fig. 1b).

To monitor loading, trapping, and release electrically and optically, we recorded ionic current and Cy5 acceptor fluorescence emission under Cy3 donor excitation with a 532 nm laser (Fig. 1c). Upon application of a +100 mV loading bias to Device 1, we initially observed an open-pore baseline ionic current consistent with an empty nanocavity. A sudden decrease in ionic current by 1.1 nA signaled nucleosome loading into the nanocavity through the *trans* nanopore, whose size was chosen comparable to the largest nucleosome dimension of approximately 11 nm. Simultaneous Cy5 fluorescence increase confirmed the nucleosome loading. The smaller *cis* nanopore prevented translocation, effectively trapping the nucleosome within the nanocavity under bias. After removing bias, ionic current disappeared, however, Cy5 fluorescence remained, indicating entropic trapping for >30 s (Fig. 1c, Supplementary Fig. 5a and b). This trapping occurred in a completely non-perturbative manner, requiring no external force to maintain confinement. Finally, reversing the bias voltage to −100 mV could force the release of the nucleosome from the nanocavity, as confirmed in Fig. 1c and Supplementary Fig. 5b, where both the magnitude of the ionic current and the fluorescence emission intensity return to the values near the baseline. In direct contrast, only translocation events were observed with Device 2 (Supplementary Fig. 5d and e), which had two larger nanopores (23 nm *trans*, 16 nm *cis*). Furthermore, a single nucleosome loaded into Device 3, featuring a larger 21 nm *trans* nanopore but smaller 10 nm *cis* nanopore, escaped immediately upon the removal of the electrical bias (Supplementary Fig. 5f and g). Repeat experiments show the same immediate escape of the loaded molecules upon removal of the bias in this device (Supplementary Fig. 5h), confirming that surface sticking of nucleosomes was prevented by BSA passivation. All subsequent experiments were performed with Device 1.

Interestingly, we did not observe the subsequent loading of a second nucleosome into Device 1 while maintaining a constant +100 mV loading bias, even after prolonged periods (Supplementary Fig. 5c). We reason that the first loaded molecule partially blocked the *cis* nanopore, causing the applied electrical bias to be redistributed across the increased resistance in the *cis* nanopore. We hypothesize that this reduces the effective capture volume, thereby decreasing the likelihood of capturing a second nucleosome near the *trans* nanopore. To test this, we increased the loading bias to +200 mV and +300 mV. The stronger electric field facilitated the sequential loading of multiple nucleosomes within a short time, as evidenced by stepwise drops in ionic current and corresponding fluctuations in fluorescence intensity (Supplementary Fig. 6). We attribute these optical signal fluctuations to the rapid accumulation of multiple fluorophores, which exceeded the microscope's resolution.

**Controlled trapping and release of multiple nucleosomes**

To achieve controlled loading of multiple nucleosomes, we incrementally increased the loading bias voltage (Fig. 2a). Under an initial +100 mV bias, an open-pore baseline ionic current of 5.4 nA was observed (Fig. 2b). Trapping of the first nucleosome reduced the ionic current to 4.8 nA, accompanied by a marked increase in fluorescence intensity from baseline to level (i), correlating with the current decrease. Increasing the loading bias to +200 mV elevated the ionic current to 11.6 nA, consistent with the baseline level for a single trapped nucleosome. Subsequent capture of a second nucleosome decreased the current to 10.7 nA, while the fluorescence intensity rose from level (i) to level (ii), confirming the presence of two single-labelled nucleosomes within the nanocavity. Upon removal of the loading bias, the two nucleosomes remained stably trapped for ~10 seconds before one escaped, indicated by a decrease in fluorescence intensity to level (iii), matching the intensity initially observed for a single trapped nucleosome (level (ii)). Release of the remaining nucleosome was achieved by applying a reverse bias of −100 mV, as evidenced by an increase in ionic current and a further reduction in fluorescence intensity.

**The nanocavity enables single-molecule FRET measurements**

Next, we sought to demonstrate the capability of our device to enable single-molecule FRET measurements on nucleosomes confined within the nanocavities. To enhance data acquisition efficiency, we employed a $5 \times 5$ nanocavity array (Supplementary Fig. 7). However, the optical signal from each nanocavity was still analysed independently, treating each as a single nanocavity device. To validate that our nanocavity enables FRET-based measurements of relative distances on the nanometer scale, we reconstituted two types of nucleosomes. Each contained a 78-bp DNA linker extending from one side of the histone core, with a Cy3 donor fluorophore conjugated to histone H2A(K120C) and a Cy5 acceptor fluorophore positioned at the 5′ end of the short DNA linker on the opposite side. The nucleosome constructs differed in that they contained either a 3-bp or a 19-bp DNA linker, which altered the proximity of the donor and acceptor fluorophores (Fig. 3a). Single-step photobleaching events observed with nanocavity-loaded nucleosomes confirmed the presence of single Cy3 donor and Cy5 acceptor dyes (Supplementary Fig. 8). To eliminate the influence of the electric field, we measured Cy3 and Cy5 fluorescence emissions from individual nucleosomes trapped within the nanocavity after removing the voltage, with Cy3 donor excitation performed using a 532-nm laser (Fig. 3b). The mean FRET values for nucleosomes with 3-bp or 19-bp short DNA linkers differed substantially (0.45 and 0.21, respectively) (Fig. 3b), consistent with the closer proximity of donor and acceptor fluorophores in the 3-bp linker construct.

**Influence of electric field on nucleosome conformation**

Electric fields could perturb biomolecular conformation, destabilizing metastable states and reducing their lifetimes under high field strength[59,60]. FRET provides a sensitive tool to detect such conformational changes on the nanometer scale. Since our nanocavity allows the observation of single nucleosomes both with and without an applied electric field, we reason that combining the nanocavity device with smFRET analysis would enable the investigation of the electric field's impact on the conformation of trapped nucleosomes. For direct comparison, we therefore also constructed mean histograms for both nucleosome types during trapping under a maintained +100 mV bias (Fig. 3c). The mean FRET value for the 3-bp linker nucleosome remained constant at approximately 0.45, regardless of the presence or absence of the electric field (Fig. 3d). In contrast, the mean FRET value for the 19-bp linker nucleosome increased substantially, from ~0.21 in the absence of an electric field to 0.39 under +100 mV bias. This marked increase in FRET efficiency indicates that the 19-bp DNA linker bends toward the histone core under the influence of the electric field, bringing the fluorophores into closer proximity. By comparison, the 3-bp linker is too short to undergo similar deformation. The effect of the applied voltage on nucleosome conformation is challenging to predict using simple assumptions, given the highly charged nature and asymmetry of the tested complexes. These results were also consistent across different devices as displayed in Supplementary Fig. 9.

To better understand how the electric field induces bending of the 19-bp linker toward the histone core, we performed a series of all-atom molecular dynamics (MD) simulations. The configuration of the system is shown in Fig. 3e. The nucleosome is modeled near the entrance of the *cis* nanopore (green) in the setup illustrated in Fig. 1a and is simulated under an electric field corresponding to +100 mV. The simulations revealed that the reduction in fluorophore distance under the applied voltage arises from a combination of electric pulling forces acting on the longer 78-bp (39-bp in the simulation box) DNA tail and the confinement of the nucleosome at the pore (blue line in Fig. 3f and Supplementary Fig. 10). Specifically, the electric pulling force exerted on the longer DNA tail through the pore drives the nucleosome against the pore walls. This interaction, combined with the funnel-shaped entrance, causes the 19-bp DNA linker to bend around the nucleosome. In all the three MD replicas we performed, the fluorophore distance decreased as the nucleosome became trapped at the nanopore. After 150 ns, the distance stabilized, oscillating around the Förster distance (6 nm), as shown in Fig. 3f (MD mean 6.2 ± 0.4 nm). Notably, the distance distribution exhibited a peak at around 6.5 nm, corresponding to an average FRET efficiency of 0.38 (Fig. 3g). As a control, we also simulated the nucleosome without any applied voltage or confinement, within an open water box (pink line in Fig. 3f and Supplementary Fig. 10). In this case, the fluorophore distance (7.7 ± 0.4 nm) was consistently larger than the Förster distance, with a corresponding calculated FRET efficiency of 0.2 ± 0.1 (Fig. 3g). These values are in good agreement with the experimental results (Fig. 3d).

**Observing the dynamics of individual nucleosomes and weak interactions of two nucleosomes inside the nanocavity**

To evaluate the ability of our nanocavity device to monitor the dynamics of individual molecules using single-molecule FRET, we investigated nucleosomal DNA unwrapping and rewrapping dynamics induced by the Chd1 chromatin remodeler[61]. We reconstituted nucleosomes with a donor fluorophore (Cy3) attached to the long DNA linker and an acceptor fluorophore (Cy5) attached to the short DNA linker, positioning the fluorophores in close proximity (Fig. 4a). In this labelling scheme, unwrapping of the outer DNA gyre from the nucleosome increases the distance between the fluorophores, resulting in a decrease in FRET efficiency. Upon trapping nucleosomes in the nanocavity, we recorded fluorescence signals from individual complexes in the presence of Chd1 and the non-hydrolysable ATP analog ATP-γS, with no applied electric field (Fig. 4b). FRET time traces revealed repeated transitions between high- and low-FRET states, indicative of a dynamic equilibrium between fully wrapped and unwrapped DNA conformations. These observations align with the known ability of Chd1 to transiently and reversibly unwrap the outer DNA gyre from the nucleosome[62]. In addition, we reconstituted nucleosomes labelled with either Cy3 or Cy5 (Fig. 4c) and demonstrated sequential loading of a Cy3-labelled and a Cy5-labelled nucleosome from a mixed solution, along with analysis of their weak interactions in the absence of an electric field. As shown in Fig. 4d, the donor (green) intensity initially increased, confirming the loading of a Cy3-labelled nucleosome at +100 mV. Increasing the voltage bias to +200 mV subsequently loaded a Cy5-labelled nucleosome. The decrease in donor and simultaneous increase in acceptor (red) intensity indicate the stacking of two nucleosomes. Upon removal of the electric field, the FRET efficiency (black) gradually decays, showing their slow dissociation (Fig. 4c). The same dissociation trend upon removal of the electric field was also observed for a pre-stacked nucleosome complex loaded into the nanocavity (Supplementary Fig. 11a). In contrast, the FRET efficiency increased in the presence of an electric field (Supplementary Fig. 11b). Thus, our nanocavity device enables non-perturbative trapping of individual macromolecular complexes and facilitates real-time observation of their molecular dynamics.

**Observing LacI-$O_3$ operator interactions at the single-molecule level under near-physiological salt concentrations**

Sequence-specific recognition of DNA by proteins - including transcription factors, polymerases, and DNA-modifying enzymes - underpins virtually all gene regulation and genome maintenance. Inside the cell, protein-DNA interactions are often stabilized by local confinement and molecular crowding, that increase the effective local concentrations of interacting partners. Such conditions have proven difficult to replicate *in vitro*, often necessitating the use of non-physiological buffer conditions to render weak interactions observable. For example, in our earlier single-molecule FRET work we were only able to detect interactions between the transcription factor LacI and a weak, naturally occurring $O_3$ operator at extremely low, non-physiological salt concentrations, because salt screens transcription factor-DNA interactions[33]. To address this issue in this work, we preloaded and immobilized (based on biotin-streptavidin interaction) an individual, Cy5-labelled, $O_3$-containing DNA molecule inside our nanocavity (see the DNA loading results in Fig. S12). Next, we loaded a single Cy3-labelled LacI homodimer into the nanocavity from a dilute solution at 50 mM KCl to create a high effective local concentration at μM range without saturating the fluorescence readout. In our labelling scheme, the fluorophores are positioned such that binding of LacI to the operator yields a FRET signal of 0.8 or 0.2, depending on whether the Cy5-proximal or distal LacI monomer is Cy3-labelled. Doubly labelled LacI dimers were excluded from analysis. LacI binding to $O_3$ (Fig. 5), as indicated by the occurrence of FRET, was observed immediately upon loading (appearance of Cy3 fluorescence), consistent with a binding time that is much faster (<1 s) than previously observed even at low salt concentrations[33]. Notably, FRET traces from individual LacI dimers exhibited instantaneous transitions between the lower and higher FRET values, corresponding to 'flipping transitions' between the two binding orientations. These presumably arise from microscopic dissociation events, in which initially $O_3$-bound LacI transiently leaves the operator and reverts its orientation before rebinding the operator.

We thus anticipate that our nanodevice will provide a powerful tool for probing a wide range of fundamentally important weak interactions, including those in the regulation of transcription, chromatin organization, and condensate formation.

**Conclusions**

In this work, we present a nanopore-gated, sub-attoliter silicon nanocavity device with customizable pore sizes, designed for non-perturbative single-molecule trapping and analysis. To accommodate the 11 nm size of the nucleosome, our model molecule, we developed an electron beam-induced carbon deposition technique to precisely reduce the nanopore gate sizes to 12 nm and 9 nm. Using ionic current as real-time feedback, we achieved controlled loading of single or multiple nucleosomes into the nanocavity. The nucleosomes were stably trapped inside the nanocavity through entropy-driven confinement, requiring no external forces. Our single-molecule FRET analyses of donor-acceptor-labelled nucleosomes showcase the capability of the device to probe the dynamics of biologically important macromolecular complexes under defined conditions. Our data further demonstrate that an applied electric field can modulate the conformational properties of the macromolecules, emphasizing a key advantage of our device: it does not require an electric field to retain trapped molecules. Furthermore, the nanocavity devices were fabricated using standard silicon technology, enabling scalability and cost-effective mass production in conventional chip manufacturing foundries. By confining individual molecules within a sub-attoliter volume, the nanocavity achieves effective concentrations approaching 10 μM for a single molecule. By investigating how proteins bind with DNA in the nanocavity, we demonstrate the nanoscale confinement to facilitate the study of biological processes requiring high local concentrations, while also addressing traditional challenges such as signal saturation in fluorescence-based approaches. Consequently, the nanocavity will enable the investigation of weak molecular interactions at the single-molecule level that may otherwise evade detection in physiologically relevant environments. We anticipate that the throughput of our nanocavity chip will be substantially augmented *via* integration with a compatible microfluidic system capable of precise, sequential sample exchange. We envision this nanocavity platform as a powerful tool for the non-perturbative interrogation of molecular dynamics, offering a powerful means to examine weak and transient interactions central to biological regulation.

# References

1. Camunas-Soler, J., Ribezzi-Crivellari, M. & Ritort, F. Elastic Properties of Nucleic Acids by Single-Molecule Force Spectroscopy. *Annu. Rev. Biophys.* **45**, 65–84 (2016).

2. Choi, J., Grosely, R., Puglisi, E. V. & Puglisi, J. D. Expanding single-molecule fluorescence spectroscopy to capture complexity in biology. *Current Opinion in Structural Biology* **58**, 233–240 (2019).

3. Dangkulwanich, M., Ishibashi, T., Bintu, L. & Bustamante, C. Molecular Mechanisms of Transcription through Single-Molecule Experiments. *Chem. Rev.* **114**, 3203–3223 (2014).

4. Schmid, S. & Dekker, C. Nanopores: a versatile tool to study protein dynamics. *Essays in Biochemistry* **65**, 93–107 (2021).

5. Hoskins, A. A., Gelles, J. & Moore, M. J. New insights into the spliceosome by single molecule fluorescence microscopy. *Current Opinion in Chemical Biology* **15**, 864–870 (2011).

6. De Vlaminck, I. & Dekker, C. Recent Advances in Magnetic Tweezers. *Annu. Rev. Biophys.* **41**, 453–472 (2012).

7. Bacic, L., Sabantsev, A. & Deindl, S. Recent advances in single-molecule fluorescence microscopy render structural biology dynamic. *Current Opinion in Structural Biology* **65**, 61–68 (2020).

8. Mohapatra, S., Lin, C.-T., Feng, X. A., Basu, A. & Ha, T. Single-Molecule Analysis and Engineering of DNA Motors. *Chem. Rev.* **120**, 36–78 (2020).

9. Felce, J. H., Davis, S. J. & Klenerman, D. Single-Molecule Analysis of G Protein-Coupled Receptor Stoichiometry: Approaches and Limitations. *Trends in Pharmacological Sciences* **39**, 96–108 (2018).

10. Bai, L., Santangelo, T. J. & Wang, M. D. SINGLE-MOLECULE ANALYSIS OF RNA POLYMERASE TRANSCRIPTION. *Annu. Rev. Biophys. Biomol. Struct.* **35**, 343–360 (2006).

11. Stracy, M. & Kapanidis, A. N. Single-molecule and super-resolution imaging of transcription in living bacteria. *Methods* **120**, 103–114 (2017).

12. Orrit, M., Ha, T. & Sandoghdar, V. Single-molecule optical spectroscopy. *Chem. Soc. Rev.* **43**, 973 (2014).

13. Kulzer, F. & Orrit, M. SINGLE-MOLECULE OPTICS. *Annu. Rev. Phys. Chem.* **55**, 585–611 (2004).

14. Michaelis, J. & Treutlein, B. Single-Molecule Studies of RNA Polymerases. *Chem. Rev.* **113**, 8377–8399 (2013).

15. Zhou, J., Schweikhard, V. & Block, S. M. Single-molecule studies of RNAPII elongation. *Biochimica et Biophysica Acta (BBA) - Gene Regulatory Mechanisms* **1829**, 29–38 (2013).

16. Juette, M. F. *et al.* The bright future of single-molecule fluorescence imaging. *Current Opinion in Chemical Biology* **20**, 103–111 (2014).

17. Hill, F. R., Monachino, E. & Van Oijen, A. M. The more the merrier: high-throughput single-molecule techniques. *Biochemical Society Transactions* **45**, 759–769 (2017).

18. Lerner, E. *et al.* Toward dynamic structural biology: Two decades of single-molecule Förster resonance energy transfer. *Science* **359**, eaan1133 (2018).

19. Dulin, D., Berghuis, B. A., Depken, M. & Dekker, N. H. Untangling reaction pathways through modern approaches to high-throughput single-molecule force-spectroscopy experiments. *Current Opinion in Structural Biology* **34**, 116–122 (2015).

20. Banterle, N. & Lemke, E. A. Nanoscale devices for linkerless long-term single-molecule observation. *Current Opinion in Biotechnology* **39**, 105–112 (2016).

21. Tyagi, S. *et al.* Continuous throughput and long-term observation of single-molecule FRET without immobilization. *Nat Methods* **11**, 297–300 (2014).

22. Kamagata, K. *et al.* Long-Term Observation of Fluorescence of Free Single Molecules To Explore Protein-Folding Energy Landscapes. *J. Am. Chem. Soc.* **134**, 11525–11532 (2012).

23. Wilson, H. & Wang, Q. ABEL-FRET: tether-free single-molecule FRET with hydrodynamic profiling. *Nat Methods* **18**, 816–820 (2021).

24. Ejaz, A., Vaidya, K. & Squires, A. H. Dynamic excitation for photophysical control and enhanced FRET sensing in an Anti-Brownian ELectrokinetic (ABEL) Trap. in *Optical Trapping and Optical*

*Micromanipulation XX* (eds Dholakia, K. & Spalding, G. C.) 106 (SPIE, San Diego, United States, 2023). doi:10.1117/12.2680788.

25. Mao, D. *et al.* Cubic DNA nanocage-based three-dimensional molecular beacon for accurate detection of exosomal miRNAs in confined spaces. *Biosensors and Bioelectronics* **204**, 114077 (2022).

26. Sengupta, B. *et al.* The Effects of Histone H2B Ubiquitylations on the Nucleosome Structure and Internucleosomal Interactions. *Biochemistry* **61**, 2198–2205 (2022).

27. Bustamante, C. J., Chemla, Y. R., Liu, S. & Wang, M. D. Optical tweezers in single-molecule biophysics. *Nat Rev Methods Primers* **1**, 25 (2021).

28. Aguirre Rivera, J. *et al.* Massively parallel analysis of single-molecule dynamics on next-generation sequencing chips. *Science* **385**, 892–898 (2024).

29. Severins, I. *et al.* Single-molecule structural and kinetic studies across sequence space. *Science* **385**, 898–904 (2024).

30. Du, J., Yin, H., Lu, Y., Lu, T. & Chen, T. Effects of Surface Tethering on the Thermodynamics and Kinetics of Frustrated Protein Folding. *J. Phys. Chem. B* **126**, 4776–4786 (2022).

31. Zou, X. *et al.* Investigating the Effect of Two-Point Surface Attachment on Enzyme Stability and Activity. *J. Am. Chem. Soc.* **140**, 16560–16569 (2018).

32. Hoarau, M., Badieyan, S. & Marsh, E. N. G. Immobilized enzymes: understanding enzyme – surface interactions at the molecular level. *Org. Biomol. Chem.* **15**, 9539–9551 (2017).

33. Marklund, E. *et al.* DNA surface exploration and operator bypassing during target search. *Nature* **583**, 858–861 (2020).

34. Schmid, S., Stömmer, P., Dietz, H. & Dekker, C. Nanopore electro-osmotic trap for the label-free study of single proteins and their conformations. *Nat. Nanotechnol.* **16**, 1244–1250 (2021).

35. Eberle, P. *et al.* Single entity resolution valving of nanoscopic species in liquids. *Nature Nanotech* **13**, 578–582 (2018).

36. Cohen, A. E. & Moerner, W. E. Method for trapping and manipulating nanoscale objects in solution. *Applied Physics Letters* **86**, 093109 (2005).

37. Chu, J. *et al.* Single-molecule fluorescence multiplexing by multi-parameter spectroscopic detection of nanostructured FRET labels. *Nat. Nanotechnol.* **19**, 1150–1157 (2024).

38. Wang, Q., Goldsmith, R. H., Jiang, Y., Bockenhauer, S. D. & Moerner, W. E. Probing Single Biomolecules in Solution Using the Anti-Brownian Electrokinetic (ABEL) Trap. *Acc. Chem. Res.* **45**, 1955–1964 (2012).

39. Cohen, A. E. & Moerner, W. E. Suppressing Brownian motion of individual biomolecules in solution. *Proc. Natl. Acad. Sci. U.S.A.* **103**, 4362–4365 (2006).

40. Liu, W. *et al.* The Electric Field in Solid State Nanopores Causes Dissociation of Strong Biomolecular Interactions. *Nano Lett.* acs.nanolett.5c01447 (2025) doi:10.1021/acs.nanolett.5c01447.

41. Chafai, D. E. *et al.* Reversible and Irreversible Modulation of Tubulin Self-Assembly by Intense Nanosecond Pulsed Electric Fields. *Advanced materials* **31**, 1903636 (2019).

42. Rodrigues, R. M., Avelar, Z., Vicente, A. A., Petersen, S. B. & Pereira, R. N. Influence of moderate electric fields in β-lactoglobulin thermal unfolding and interactions. *Food Chemistry* **304**, 125442 (2020).

43. Krishnan, M., Mojarad, N., Kukura, P. & Sandoghdar, V. Geometry-induced electrostatic trapping of nanometric objects in a fluid. *Nature* **467**, 692–695 (2010).

44. Mojarad, N. & Krishnan, M. Measuring the size and charge of single nanoscale objects in solution using an electrostatic fluidic trap. *Nature Nanotech* **7**, 448–452 (2012).

45. Ruggeri, F. *et al.* Single-molecule electrometry. *Nature Nanotech* **12**, 488–495 (2017).

46. Rosenkranz, T. *et al.* Observing Proteins as Single Molecules Encapsulated in Surface-Tethered Polymeric Nanocontainers. *ChemBioChem* **10**, 702–709 (2009).

47. Cisse, I., Okumus, B., Joo, C. & Ha, T. Fueling protein–DNA interactions inside porous nanocontainers. *Proc. Natl. Acad. Sci. U.S.A.* **104**, 12646–12650 (2007).

48. Liu, X., Skanata, M. M. & Stein, D. Entropic cages for trapping DNA near a nanopore. *Nat Commun* **6**, 6222 (2015).

49. Lam, M. H. *et al.* Entropic Trapping of DNA with a Nanofiltered Nanopore. *ACS Appl. Nano Mater.* **2**, 4773–4781 (2019).

50. Gao, C. *et al.* Effects of Molecular Crowding on the Structure, Stability, and Interaction with Ligands of G-quadruplexes. *ACS Omega* **8**, 14342–14348 (2023).

51. Hirai, M., Arai, S. & Iwase, H. Fibrillization Process of Human Amyloid-Beta Protein (1–40) under a Molecular Crowding Environment Mimicking the Interior of Living Cells Using Cell Debris. *Molecules* **28**, 6555 (2023).

52. Zhuang, X. *et al.* A Single-Molecule Study of RNA Catalysis and Folding. *Science* **288**, 2048–2051 (2000).

53. Ha, T. *et al.* Probing the interaction between two single molecules: fluorescence resonance energy transfer between a single donor and a single acceptor. *Proc. Natl. Acad. Sci. U.S.A.* **93**, 6264–6268 (1996).

54. Ha, T. Single-Molecule Fluorescence Resonance Energy Transfer. *Methods* **25**, 78–86 (2001).

55. Zeng, S., Chinappi, M., Cecconi, F., Odijk, T. & Zhang, Z. DNA compaction and dynamic observation in a nanopore gated sub-attoliter silicon nanocavity. *Nanoscale* **14**, 12038–12047 (2022).

56. Van Dorp, S., Keyser, U. F., Dekker, N. H., Dekker, C. & Lemay, S. G. Origin of the electrophoretic force on DNA in solid-state nanopores. *Nature Phys* **5**, 347–351 (2009).

57. Gubbiotti, A. *et al.* Electroosmosis in nanopores: computational methods and technological applications. *Advances in Physics: X* **7**, 2036638 (2022).

58. Luger, K. Crystal structure of the nucleosome core particle at 2.8 A˚ resolution. **389**, (1997).

59. Bekard, I. & Dunstan, D. E. Electric field induced changes in protein conformation. *Soft Matter* **10**, 431–437 (2014).

60. Liu, S.-C., Ying, Y.-L., Li, W.-H., Wan, Y.-J. & Long, Y.-T. Snapshotting the transient conformations and tracing the multiple pathways of single peptide folding using a solid-state nanopore. *Chem. Sci.* **12**, 3282–3289 (2021).

61. Bowman, G. D. & Deindl, S. Remodeling the genome with DNA twists. *Science* **366**, 35–36 (2019).

62. Farnung, L., Vos, S. M., Wigge, C. & Cramer, P. Nucleosome–Chd1 structure and implications for chromatin remodelling. *Nature* **550**, 539–542 (2017).

63. Zeng, S. *et al.* Controlled size reduction and its underlying mechanism to form solid-state nanopores via electron beam induced carbon deposition. *Nanotechnology* **30**, 455303 (2019).

64. Deindl, S. & Zhuang, X. Monitoring conformational dynamics with single-molecule fluorescence energy transfer: applications in nucleosome remodeling. in *Methods in enzymology* vol. 513 59–86 (Elsevier, 2012).

65. Sabantsev, A. *et al.* Spatiotemporally controlled generation of NTPs for single-molecule studies. *Nat Chem Biol* **18**, 1144–1151 (2022).

66. Deindl, S. *et al.* ISWI Remodelers Slide Nucleosomes with Coordinated Multi-Base-Pair Entry Steps and Single-Base-Pair Exit Steps. *Cell* **152**, 442–452 (2013).

67. Levendosky, R. F., Sabantsev, A., Deindl, S. & Bowman, G. D. The Chd1 chromatin remodeler shifts hexasomes unidirectionally. *eLife* **5**, e21356 (2016).

68. Marklund, E. *et al.* Sequence specificity in DNA binding is mainly governed by association. *Science* **375**, 442–445 (2022).

69. Sabantsev, A., Levendosky, R. F., Zhuang, X., Bowman, G. D. & Deindl, S. Direct observation of coordinated DNA movements on the nucleosome during chromatin remodelling. *Nat Commun* **10**, 1720 (2019).

70. Deindl, S. & Zhuang, X. Monitoring Conformational Dynamics with Single-Molecule Fluorescence Energy Transfer: Applications in Nucleosome Remodeling. in *Methods in Enzymology* vol. 513 59–86 (Elsevier, 2012).

71. Plesa, C. & Dekker, C. Data analysis methods for solid-state nanopores. *Nanotechnology* **26**, 084003 (2015).

72. Schindelin, J. *et al.* Fiji: an open-source platform for biological-image analysis. *Nat Methods* **9**, 676–682 (2012).

73. Schneider, C. A., Rasband, W. S. & Eliceiri, K. W. NIH Image to ImageJ: 25 years of image analysis. *Nat Methods* **9**, 671–675 (2012).

74. Gumbart, J., Khalili-Araghi, F., Sotomayor, M. & Roux, B. Constant electric field simulations of the membrane potential illustrated with simple systems. *Biochimica et Biophysica Acta (BBA) - Biomembranes* **1818**, 294–302 (2012).

75. Baldelli, M. *et al.* Controlling electroosmosis in nanopores without altering the nanopore sensing region. *Advanced Materials* **36**, 2401761 (2024).

## Methods

### Nanopore-gated nanocavity fabrication

The fabrication process of the nanopore-gated nanocavity device began with a double-side-polished (100) silicon-on-insulator (SOI) wafer. The SOI wafer consisted of an 88-nm-thick top silicon (Si) layer, a buried 145-nm-thick oxide (BOX) layer, and a 300-μm-thick Si substrate. First, a 30-nm-thick low-stress silicon nitride ($SiN_x$) layer was deposited on both sides of the SOI wafer using low-pressure chemical vapor deposition (Koyo Lindberg). Photolithography (Karl Süss) and reactive ion etching (RIE; Advanced Vacuum) were used to open square windows (150 μm per side) on the SiN layer from the bottom side of the wafer. Subsequently, large cavities in the Si substrate were etched using a combination of deep RIE and wet etching in 80 °C KOH. During this step, the top $SiN_x$ layer was protected, and the BOX layer served as an effective etch stop for the KOH etching. Next, a 50-nm-thick gold layer was deposited onto the top $SiN_x$ layer using an evaporator (Kurt J. Lesker Company) and lifted off, leaving an uncovered $SiN_x$ region aligned with the backside window. Nanoscale holes were patterned in the uncovered top $SiN_x$ layer *via* electron beam lithography (Nanobeam Ltd) followed by RIE. Nanocavities were then formed by etching the top Si layer in 60 °C KOH through the nanoscaled holes and stripping the BOX layer. The lateral etching of the (111) Si planes primarily controlled the formation of the nanocavities. Finally, the fabricated chips were mounted on adhesive carbon tabs and loaded into a scanning electron microscope (SEM; Zeiss 1530, Germany) chamber for electron beam-induced carbon deposition[63].

### Preparation of fluorophore-labelled nucleosomes and TOTO-1 labelled DNA

Recombinant histones (H2A, H2A(K120C), H2B, H3 C110A and H4) from Xenopus laevis were purchased from the Histone Source Protein Expression and Purification Facility, Colorado State University, Fort Collins, CO. H2A(K120C) was labelled with Cy3 as described previously[64]. Briefly, one milligram of lyophilized H2A120C was diluted in unfolding buffer (20 mM Tris pH 7.0, 7 M guanidine-HCl, 5 mM EDTA, 1.25 mM TCEP) and incubated for 2 h at room temperature in the dark. Cy3-maleimide was dissolved in DMSO and added to the protein at a final concentration of 0.75 mM. After 3 h in the dark at room temperature, the reaction was quenched with a final concentration of 80 mM β-mercaptoethanol. The labelled protein was dialyzed nine times against dialysis buffer (20 mM Tris pH 7.0, 7 M guanidine-HCl, 1 mM DTT) and then used in histone dimer assembly. The labelling efficiency was approximately 70–85%.

Biotinylated and fluorescently-labelled DNA constructs containing the Widom 601 nucleosome positioning sequence were generated by PCR (for +3 and +19 nucleosomes) or by annealing and ligating a set of overlapping oligonucleotides (unwrapping nucleosomes)[65]. These DNA constructs have a long (78-bp) stretch of flanking DNA with biotin at the end on one side of the Widom 601 nucleosome positioning sequence and a short stretch of either 3 of 19 bp of flanking DNA with a Cy5 fluorophore at the end on the other side of the nucleosome positioning sequence. The unwrapping construct additionally has an internal Cy3 DNA label on the long-linker side located 4 bp away from the edge of the nucleosome positioning sequence. "Cy3-only" and "Cy5-only" DNA constructs used to monitor nucleosome dimerization have a 63-bp linker on one side and no linker on the other side of the Widom 601 nucleosome positioning sequence, with either a Cy3 or a Cy5 fluorophore attached to the 5' DNA terminus on the "no-linker" side, and were generated by PCR.

Unwrapping, "Cy3-only" and "Cy5-only" nucleosomes were directly reconstituted by salt gradient dialysis using unlabelled histone octamer and purified by preparative polyacrylamide gel electrophoresis using the Mini Prep Cell apparatus (Bio-Rad), as described previously[65]. "Cy3-Cy5-both" nucleosomes were assembled from the "Cy5-only" DNA using the H2A120C-Cy3 histone octamer. +3 and +19 nucleosomes[66] were reconstituted by adding wild-type H2A/H2B dimer to oriented hexasomes assembled with H2A120C-Cy3 as described previously[67].

20 kb DNA ladders were purchased from Fisher Scientific. TOTO-1, an intercalating fluorescent DNA dye, was used to label the DNA molecules with a nucleotide to dye ratio of 10:1.

The imaging buffer contained 20 mM Tris pH 7.0, 100 mM KCl, 0.1 mg/mL acetylated BSA (Promega), 10% (v/v) glycerol, 10% (w/v) glucose, supplemented with 2 mM Trolox to reduce photoblinking of the dyes, as well as an enzymatic oxygen scavenging system (composed of 800 μg/ml glucose oxidase and 50 μg/ml catalase). Nucleosomes were then dispersed in the imaging buffer to 1 nM and DNA to 100 pM, respectively. To study the dynamics of individual nucleosomes, 0.4 μM Chd1, 1 mM $MgCl_2$ and 1mM ATP-γS were added together with unwrapping nucleosomes. To monitor the dynamics of nucleosome dimerization, Cy3-only nucleosomes, Cy5-only nucleosomes and 1 mM $MgCl_2$ were mixed together.

**Preparation of LacI, DNA constructs**

LacI-Cy3 was purified and labelled as described previously[33,68]. Briefly, the protein contains a C-terminal 6xHis-tag for affinity purification, and lacks the C-terminal tetramerization domain, as well as all endogenous cysteines except for the one surface-inaccessible residue in the dimerization interface that is critical for dimerization. An additional cysteine residue is introduced in position 28 for fluorophore labelling (Supplemantary Table 1). LacI was expressed in TOP10 *E. coli* at 30°C for 4 h and purified on a HisTrap column and a Superdex 200 size exclusion column. Labelling of LacI with Cy3-maleimide was performed in a buffer containing 100 mM potassium phosphate pH 7.0, 150 mM NaCl, and 5 mM EDTA. 10 μM LacI and 100 μM TCEP were incubated for 30 min at room temperature in the dark. Cy3 maleimide was added in a 64-fold molar excess and the labelling reaction was incubated at room temperature for 2 h, followed by an overnight incubation at 4°C. Finally, the reaction was quenched by the addition of β-mercaptoethanol to a final concentration of 4.5 mM and excess unreacted dye was removed using Pierce Dye Removal Resin following the manufacturer's instructions.

Double-stranded DNA construct that contains an $O_3$ operator site, FRET acceptor (backbone-incorporated Cy5), and a biotin moiety at the 5' end distal to the operator, was generated by annealing two complementary oligonucleotides (IDT DNA, Supplementary Table 1). High-performance liquid chromatography (HPLC)-purified oligonucleotides were mixed at equimolar concentrations (1 μM) in 50 mM Tris pH 8.0, 100 mM KCl, 1 mM EDTA, and annealed with a temperature ramp (95–3°C over 13 hours).

**Preparation of PEG-passivated and streptavidin-modified nanocavity**

PEG-coated nanocavity was prepared according to previously published methods[69,70]. Briefly, silicon chips containing nanocavities were amino-functionalized by incubation with 1% (w/v) Vectabond (Vector Laboratories) in > 99.9% (w/v) acetone for 1 h. The chips were then treated with a mixture of 20% (w/v) PEG-SVA (MW 5000) and 0.2% biotin-PEG-SVA (MW 5000) in 0.1 M sodium bicarbonate for 4 h. After PEGylation, the chip was subsequently mounted on a flow cell, and streptavidin (3.3 μM, 100 μL) was added to both reservoirs, followed by incubation under an applied voltage of +100 mV for 1 h.

**Electrical and optical characterizations**

Prior to the electrical measurements, the nanopore chip was carefully cleaned in oxygen plasma at 1,000 W for 10 min, followed by immersion in a piranha solution with $H_2SO_4:H_2O = 3:1$ (volume ratio) for 30 min and finally rinsed in deionized water. The nanopore chip was then mounted on a custom-made polymethyl methacrylate flow cell (Supplementary Fig. 13) and the *cis* chamber was sealed using a 0.17 mm thick cover glass. Imaging buffer containing 0.1 mg/mL BSA was then added to both chambers, and the chip was immersed to passivate the solid surface. Two compartments were separated by the chip and the only path of ionic current was through the nanopore. The flow cell was placed on a widefield fluorescence microscope (Ti Eclipse, Nikon) equipped with a 63x water-immersion objective (Nikon). A pair of Ag/AgCl electrodes (2 mm in diameter (Warner Instruments LLC.)) was used to apply a bias voltage across the nanopore and to measure the ionic current. The electrical measurement was monitored using a patch clamp amplifier (Multiclamp 700B, Molecular Device Inc.). The voltage control and current record were achieved with a custom LABVIEW programme. The ionic current was digitalized by an Axon Digidata 1550B1 (Molecular Device LLC.) and recorded using the software Axon pCLAMP 11 (Molecular Device LLC.). The traces were sampled at 10 kHz and low-pass filtered at 5 kHz. Extraction of translocation events was performed with the Transalyzer package[71]. Prior to introducing the actual biological sample, we conducted pre-experiments by sweeping the I–V curve multiple times to confirm the reproducibility of regular ionic signals under episodic voltage applied across the nanocavity. Then the *trans* chamber was replaced with imaging buffer containing target molecules while the *cis* chamber only filled with imaging buffer. TOTO-1 and Cy3 fluorophores were excited with a PE-300 white broad spectrum LED illuminator (CoolLED) through a Cy3 excitation filter. Fluorescence emission in the Cy3 (or TOTO-1) and Cy5 spectral channels is projected side-by-side onto an EMCCD camera (iXon 888 Ultra, Andor). Images were processed and analysed by using Fiji/ImageJ software[72,73].

**Molecular dynamics simulation**

The simulation box is composed of a neutral solid-state membrane (green) containing a drilled nanopore and a reduced version of the nucleosome used in the experimental setup. The nucleosome model includes histones (transparent blue), 147-bp of double-stranded DNA (dsDNA) wrapped around the histone, and two elongated dsDNA tails: a shorter tail of 19-bp and a longer tail of 39-bp. The residues where the two fluorophores are attached, Cy3 (green, located on H2A K120) and Cy5 (red, at the DNA 5' end), are highlighted. To reduce computational cost, the modeled membrane thickness and the length of the longer DNA tail are halved in comparison to the experimental values (30 nm and 78-bp, respectively). Initially, the nucleosome is positioned at the nanopore entrance. After equilibration, a constant, homogeneous electric field is applied along the z-axis, $E=(0,0,E_z)$, simulating an applied voltage of +100 mV, as done in other works[74,75]. The negatively charged long tail is electrophoretically pulled through the nanopore. After 150 ns, the nucleosome reaches a stable state, with the center of mass of the protein core located between 3-5 nm above the membrane surface (Supplementary Fig. 10d).

## Acknowledgements

This work was financially supported the Wallenberg Academy Fellow Extension Programme (2020-0190 to Z.Z.) and partially supported by the Olle Engkvist Foundation (214-0322 to Z.Z.), the Swedish Strategic Research Foundation (FFL15-0174 to Z.Z.) as well as the European Research Council (ERC) Advanced Grant ERC-ADG-101092623 (S.D.), the Knut and Alice Wallenberg Foundation grants KAW/WAF 2019.0306 and KAW 2024.0012 (S.D.), Cancerfonden grant 25 4453 Pj (S.D.), and Swedish Research Council project grant VR 03255 (S.D.); For Molecular Dynamics simulation this research used the HPC computational resource provided by CINECA on LEONARDO BOOSTER (projects IsB27_NOVOBAR). The device fabrication was done in the Ångström Microstructure Laboratory (MSL), Uppsala University and the technical staff of MSL are acknowledged for their process support.

## Author contributions

Z.Z. conceived the idea and initiated the project. F.L., Q.H., A.S., S.D. and Z.Z. designed the experiments. Q.H. and S.Z. initiated the device fabrication. F.L. completed device fabrication & characterization (under the supervision of S.Z. and Z.Z.), and single molecule data analysis (under the supervision of A.S and S.D). G.D.M. and M.C. performed the molecular dynamics and PNP simulations. F.L., G.D.M., A.S., S.D. and Z.Z. co-wrote the manuscript. All the authors analysed the data, discussed the results and commented on the manuscript.

## Competing interests

The authors declare no competing financial interests.

## Additional information

Supplementary Information is available for this paper. Correspondence and requests for materials should be addressed to S.D. and Z.Z.

## Data availability

The authors declare that the data supporting the findings of this study are available within the paper and its supplementary information files.

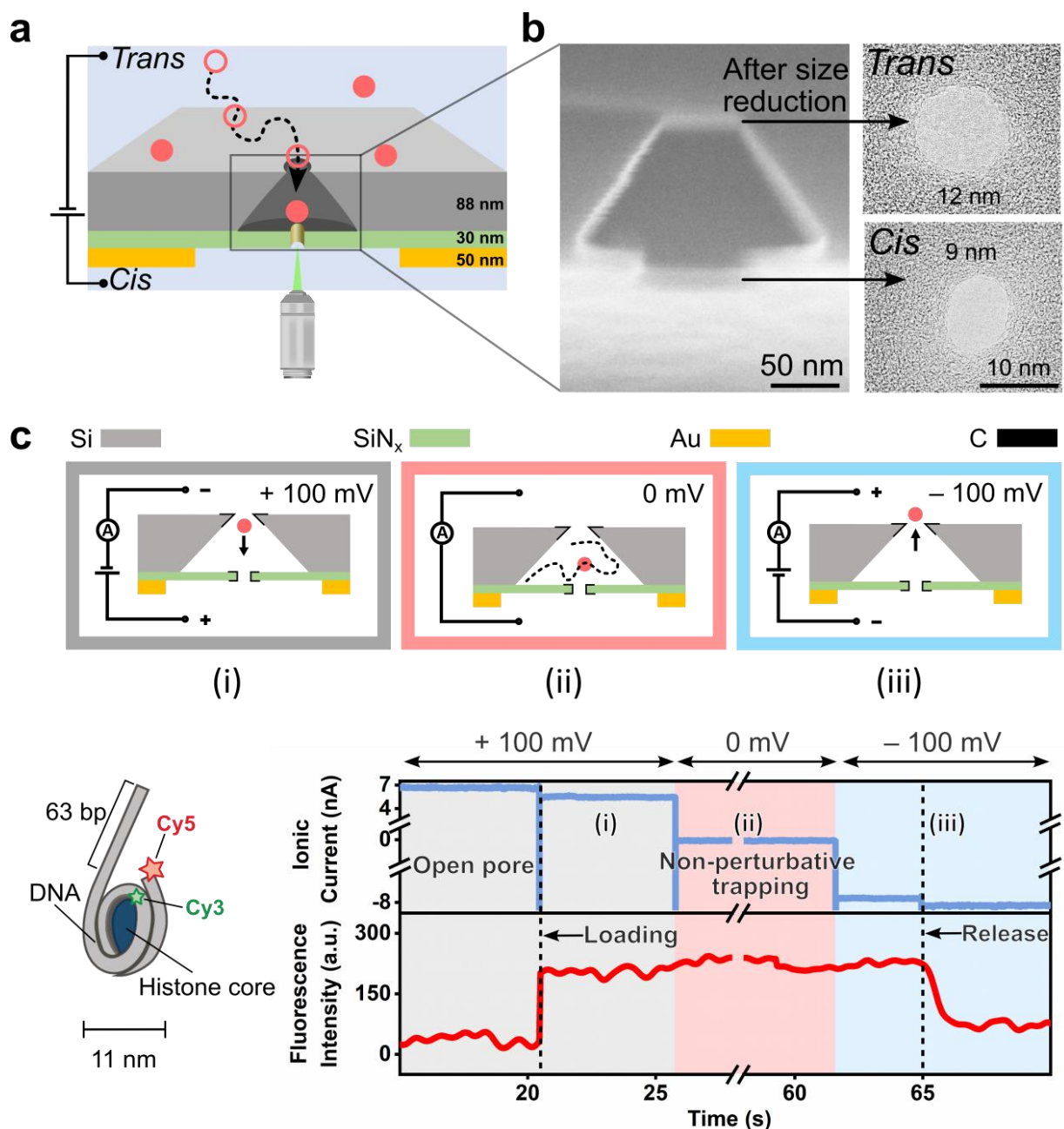

**Fig. 1 | Nanocavity device structure, experimental setup, and working principle for the controllable loading, trapping and release of a single molecule.** (a) Schematic of the nanocavity device and the electro-optical setup, with a single-molecule fluorescence confocal microscope positioned beneath the *cis* side of the nanocavity. Upon application of an electric field, individual molecules are drawn into the nanocavity. (b) Cross-sectional SEM image of the nanocavity device prior to carbon deposition, and TEM images of the silicon pore on the *trans* side and the SiN$_x$ pore on the *cis* side after carbon deposition for size reduction. (c) Upper: schematic representation of the controllable loading, trapping and release of a single molecule. (i) Loading into the nanocavity at +100 mV. (ii) Non-perturbative trapping at 0 mV, with the dashed line indicating Brownian motion. (iii) Release upon voltage reversal to −100 mV. Lower: cartoon representation of the 11-nm fluorophore-labelled nucleosome. Time traces of ionic current (blue) and Cy5 fluorescence intensity (red), corresponding to the steps described in the schematic representation, for the loading experiments of 1 nM fluorophore-labelled nucleosomes in imaging buffer.

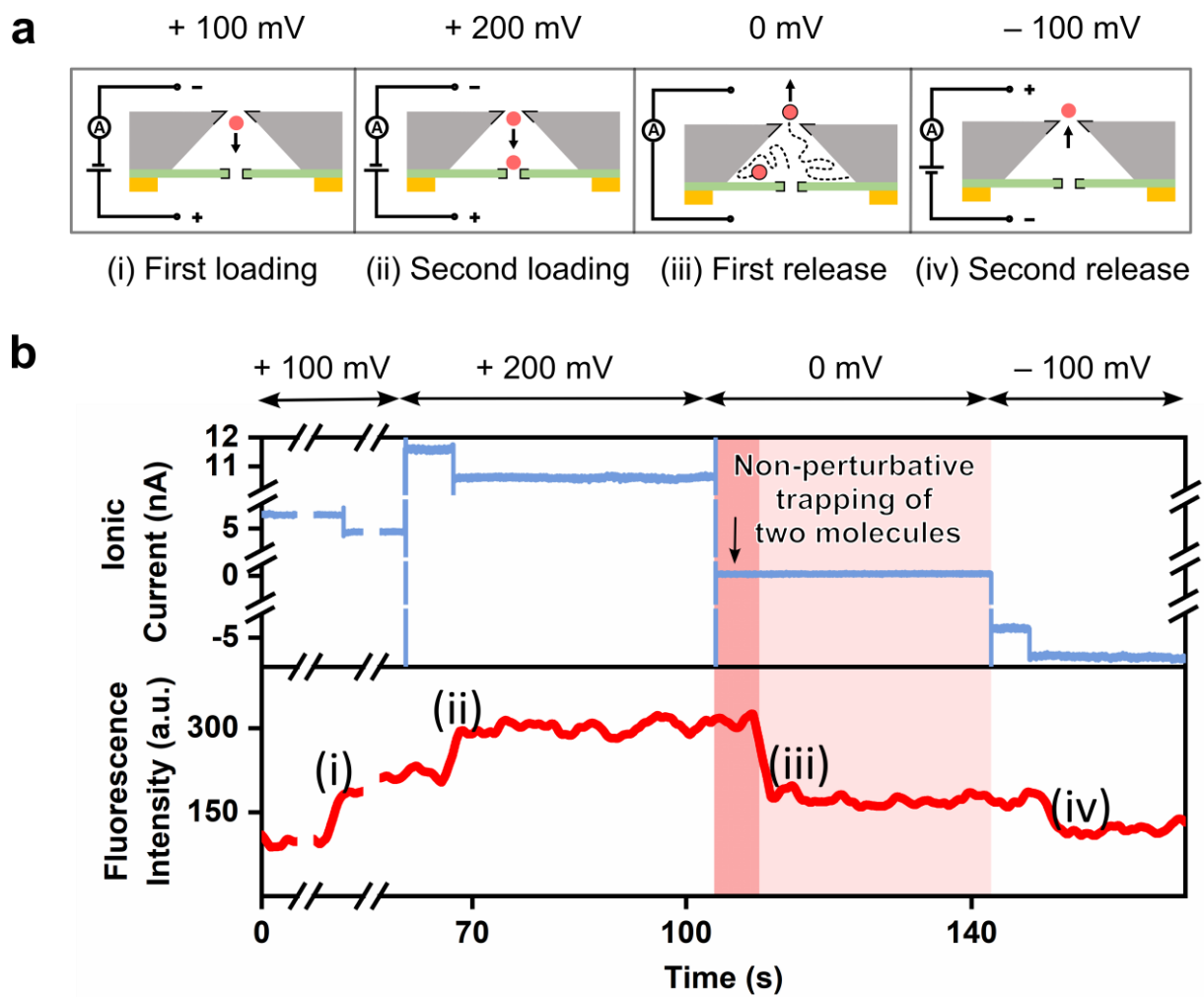

**Fig. 2 | Controlled trapping and release of multiple nucleosomes using a stepwise voltage ramp.** (a) Schematic representation of the sequential loading and release process. (i) The first nucleosome is trapped under a +100 mV bias. (ii) Increasing the voltage to +200 mV enables the capture of a second nucleosome. (iii) Removal of the voltage creates a non-perturbative confinement, during which one nucleosome escapes while the other remains trapped. (iv) Applying a reverse bias of −100 mV releases the remaining nucleosome. (b) Time traces of ionic current (blue) and Cy5 fluorescence intensity (red) corresponding to the steps described in (a). Data were recorded for 1 nM fluorophore-labelled nucleosomes in imaging buffer. Dark pink shading indicates trapping of two nucleosomes, while light pink shading represents trapping of a single nucleosome.

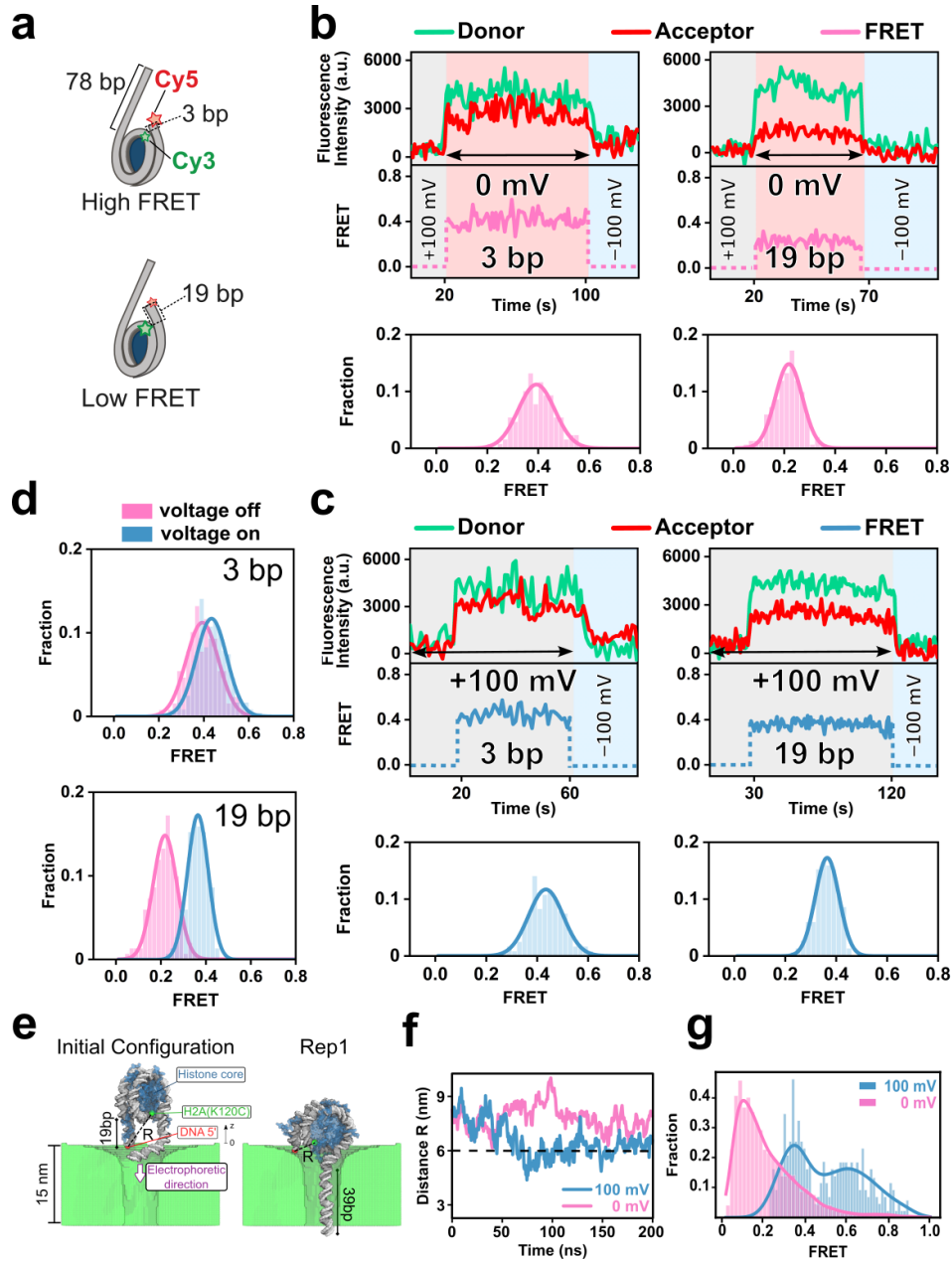

**Fig. 3 | Effect of the electric field on nucleosome conformation.** (a) Schematic representations of nucleosomes containing either a 3-bp or a 19-bp DNA linker. (b) Representative time traces of donor (green), acceptor (red), and FRET (pink) signals for 3-bp and 19-bp nucleosomes recorded at 0 mV. Histograms of the mean FRET distributions for 3-bp ($N$ = 40 events from 5 independent experiments) and 19-bp ($N$ = 45 events from 5 independent experiments) nucleosomes in the absence of voltage. Shading represents conditions: grey for +100 mV, pink for 0 mV, and blue for −100 mV. (c) Representative time traces of donor (green), acceptor (red), and FRET (blue) signals for 3-bp and 19-bp nucleosomes recorded at +100 mV. Histograms of the mean FRET distributions for 3-bp ($N$ = 43 events from 5 independent experiments) and 19-bp ($N$ = 38 events from 5 independent experiments) nucleosomes under voltage. (d) Gaussian-fitted histograms of mean FRET values for 3-bp and 19-bp nucleosomes, with and without applied voltage. (e) Molecular dynamics simulation of a nucleosome with 19-bp linker captured at the *cis* nanopore of the setup shown in Fig. 1. The Rep1 panel displays the final frame of one independent replicate under an applied electric potential +100 mV; the other two replicates, under the same conditions, are shown in Supplementary Fig. 10. (f) Time evolution of the distance (R) between the fluorophore attachment sites for replicate 1. The black dashed line marks the Förster distance ($R_0$), indicating the distance at which energy transfer is 50% efficient. Note that the fluorophores themselves are not explicitly modeled in the MD simulations, thus the measured distance reflects only the attachment site locations. Furthermore, $R_0$

depends on the relative orientation of the fluorophores. (g) FRET values derived from the distances computed in after 150 ns three MD replicates, assuming a fixed $R_0$.

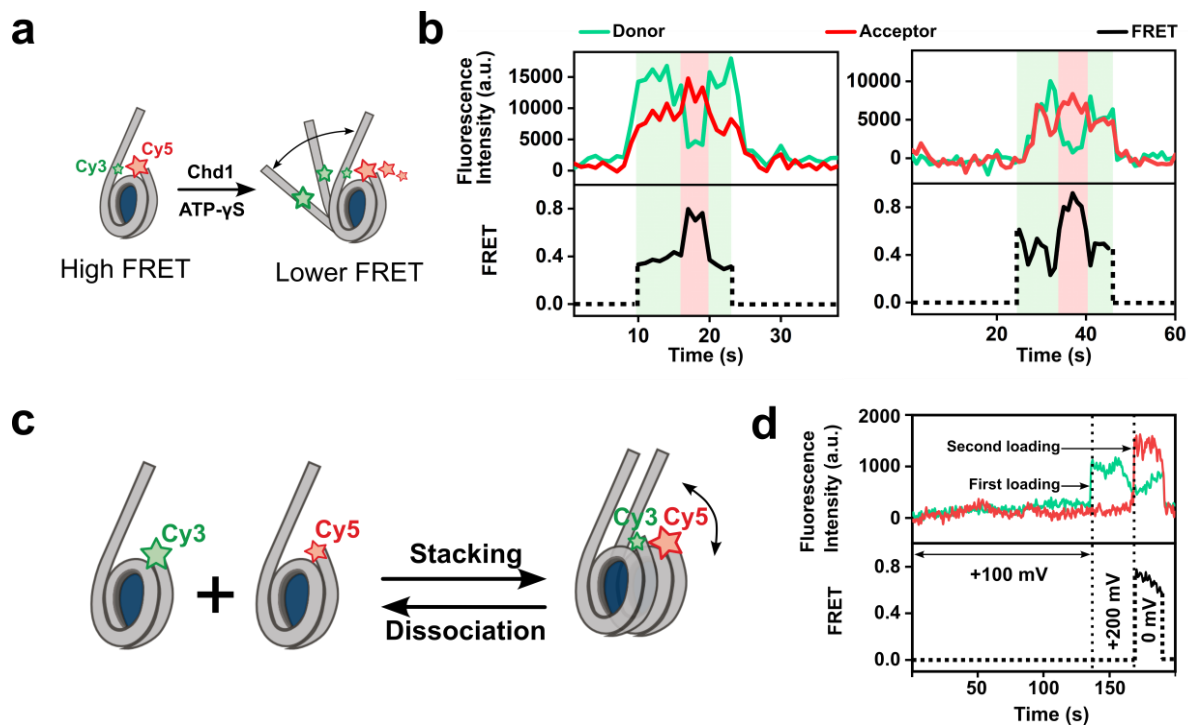

**Fig. 4 | Nucleosomal DNA unwrapping by the chromatin remodeler Chd1 and weak interaction between two nucleosomes.** (a) Schematic of the labelling scheme and the dynamic equilibrium between fully wrapped and unwrapped nucleosome conformations in the presence of Chd1 and ATP-γS. (b) Representative time traces of donor fluorescence (green), acceptor fluorescence (red), and FRET efficiency (black) illustrating Chd1-induced unwrapping and rewrapping of individual nucleosomes in the presence of ATP-γS, recorded while the nucleosomes were confined in the nanocavity at 0 mV. Shading indicates states: green for lower-FRET and red for higher-FRET states. (c) Schematic of the weak stacking interaction between two nucleosomes labelled with donor and acceptor, respectively. The conceptual cartoon is not meant to imply any particular mode of nucleosome stacking. The double-headed black arrow illustrates the dynamic nature of stacked nucleosomes: while they can form stable interactions, they undergo relative motions (e.g., tilting and rotational fluctuations). (d) Representative time traces of donor fluorescence (green), acceptor fluorescence (red), and FRET efficiency (black) illustrating the dynamics of two nucleosomes sequentially loaded from a mixed solution.

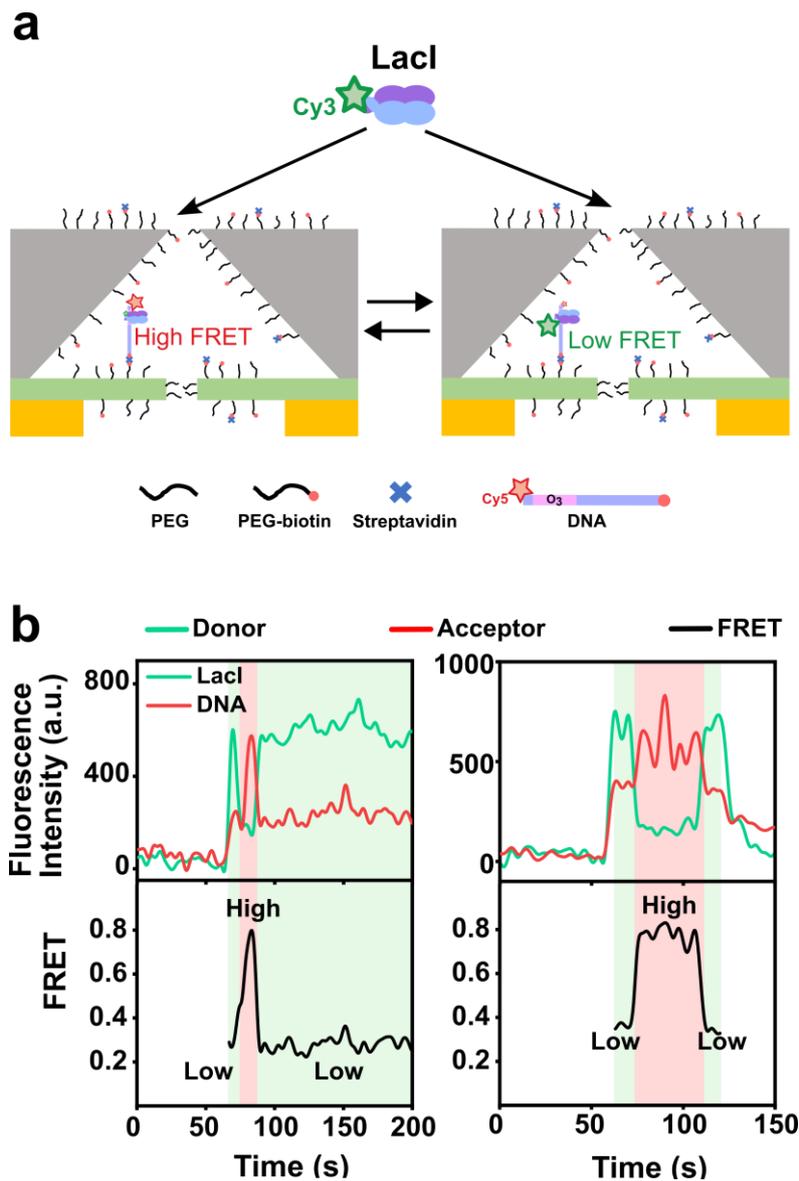

**Fig. 5 | LacI-DNA ($O_3$) dynamic interactions at near-physiological salt concentration (50 mM KCl).** (a) Schematic of the LacI loading at +100 mV and the dynamic transition between different binding orientation at 0 mV. (b) Representative time traces of LacI with donor fluorescence (green), DNA with acceptor fluorescence (red), and FRET efficiency (black) illustrating LacI exploring the DNA surface, recorded while the LacI molecules were confined in the nanocavity at 0 mV. Shading indicates states: green for lower-FRET and red for higher-FRET states.

# Supplementary Information

# A nanopore-gated sub-attoliter silicon nanocavity for single molecule trapping and analysis

**Funing Liu[1], Qitao Hu[1,6], Anton Sabantsev[2], Giovanni Di Muccio[3,4], Shuangshuang Zeng[1,7], Mauro Chinappi[5], Sebastian Deindl[2] and Zhen Zhang[1]**

[1]Division of Solid-State Electronics, Department of Electrical Engineering, Uppsala University, BOX 65, SE-75121, Uppsala, Sweden.
[2]Department of Cell and Molecular Biology, Science for Life Laboratory, Uppsala University, Uppsala, Sweden.
[3]NY-Masbic, Department of Life and Environmental Sciences, Università Politecnica delle Marche, Via Brecce Bianche, 60131 Ancona, Italy
[4]National Future Biodiversity Centre (NFBC), Palermo, Italy
[5]Department of Industrial Engineering, University of Rome Tor Vergata, Roma, Italy.
[6]Current address: Department of Radiology, Stanford University, Stanford, CA 94305 USA.
[7]Current address: School of Integrated Circuits, Huazhong University of Science and Technology, Wuhan 430074, China.

Correspondence should be addressed to S.D. (sebastian.deindl@icm.uu.se) and Z.Z. (zhen.zhang@angstrom.uu.se).

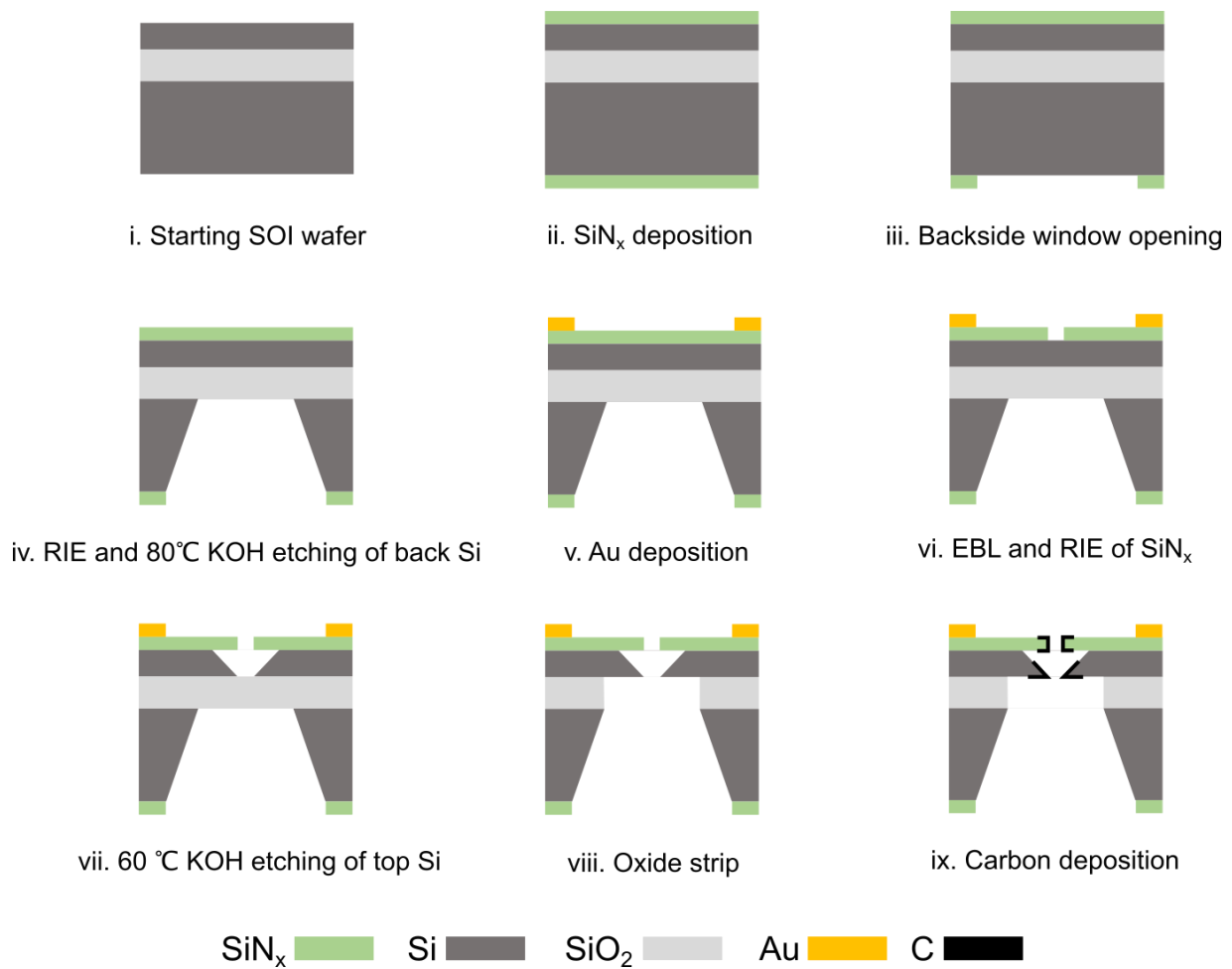

**Supplementary Fig. 1. Schematic illustration of the process flow for fabricating the nanopore-gated nanocavity device.** The fabrication process builds on our established workflow for truncated pyramidal nanopores[1,2]. (i) Starting SOI wafer. (ii) Deposition of SiNx on the SOI wafer using low-pressure chemical vapor deposition (LPCVD). (iii) Backside window opening *via* photolithography, followed by reactive ion etching (RIE). (iv) Silicon etching in the bulk substrate using deep RIE, followed by KOH wet etching at 80 °C. (v) Gold deposition (*via* metal evaporation) and lift-off on the top SiNx layer, leaving an uncovered region aligned with the backside window. (vi) Nanopore creation in the uncovered top SiNx layer using electron beam lithography (EBL) and RIE. (vii) Silicon nanocavity etching in 60 °C KOH solution. (viii) Removal of the buried oxide layer using buffered HF. (ix) Carbon deposition under SEM scanning to precisely tune the nanopore size.

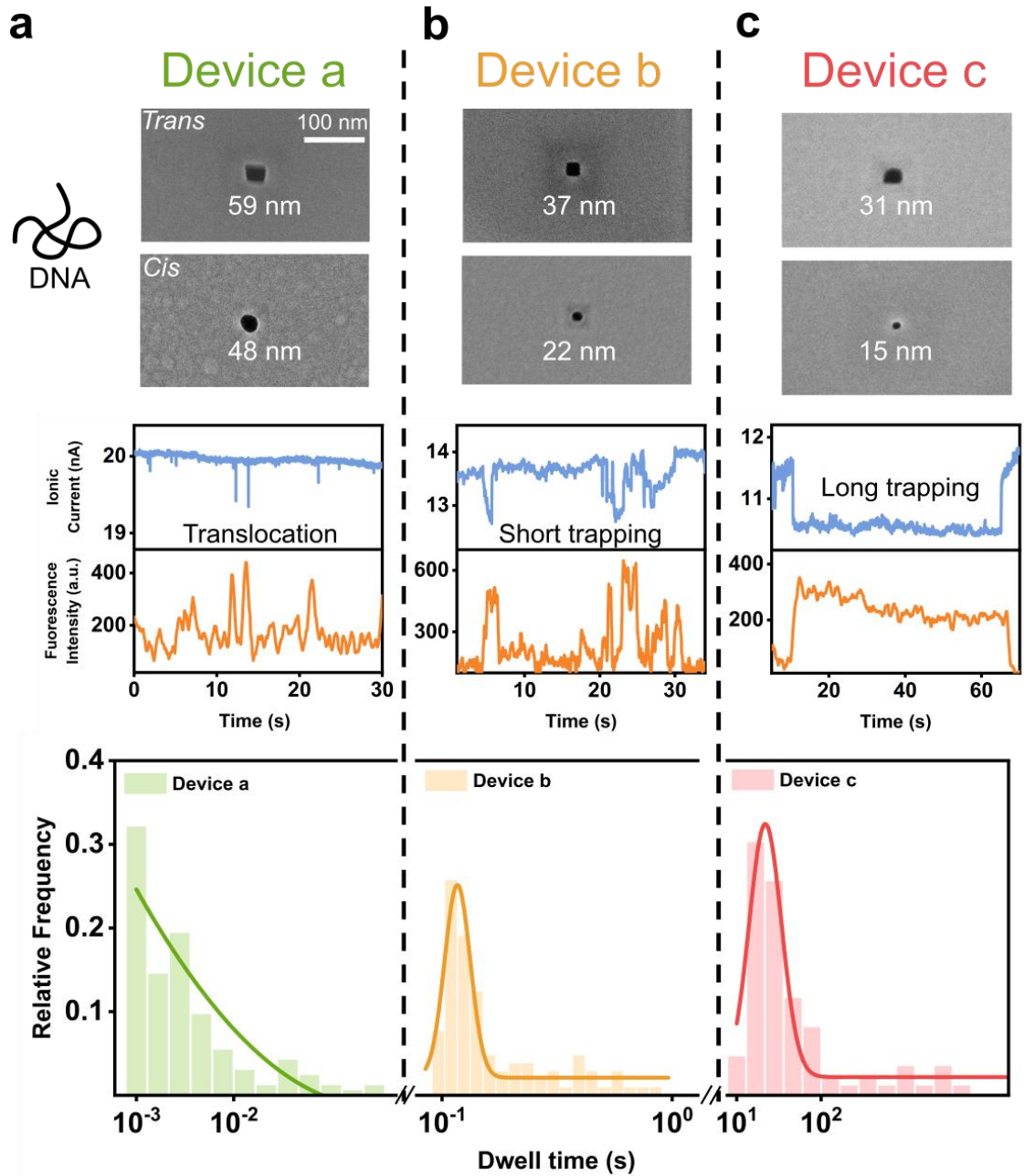

**Supplementary Fig. 2. Relationship between nanopore gate size and dwell time of 20 kb DNA in the nanopore-gated nanocavity.** Three devices with varying *trans* and *cis* nanopore sizes were fabricated to evaluate the dwell times of 20 kb DNA (hydrodynamic diameter approximately 280 nm) under +100 mV bias. Negatively charged DNA molecules in imaging buffer were captured into the nanocavity *via* electrophoretic forces. Top row: SEM images of the devices with *trans* nanopore sizes of 59 nm, 37 nm, and 31 nm and *cis* nanopore sizes of 48 nm, 22 nm, and 15 nm, respectively. Middle row: Representative ionic current (blue) and fluorescence intensity (orange) time traces for 100 pM TOTO-1-labelled 20 kb DNA at +100 mV, showing (a) rapid translocation, (b) short-term trapping, and (c) extended trapping. Bottom row: Dwell time histograms for (a) Device a (green, $N$ = 70 events from 5 independent experiments), (b) Device b (yellow, $N$ = 56 events from 5 independent experiments), and (c) Device c (red, $N$ = 32 events from 5 independent experiments), fitted with Gaussian distributions.

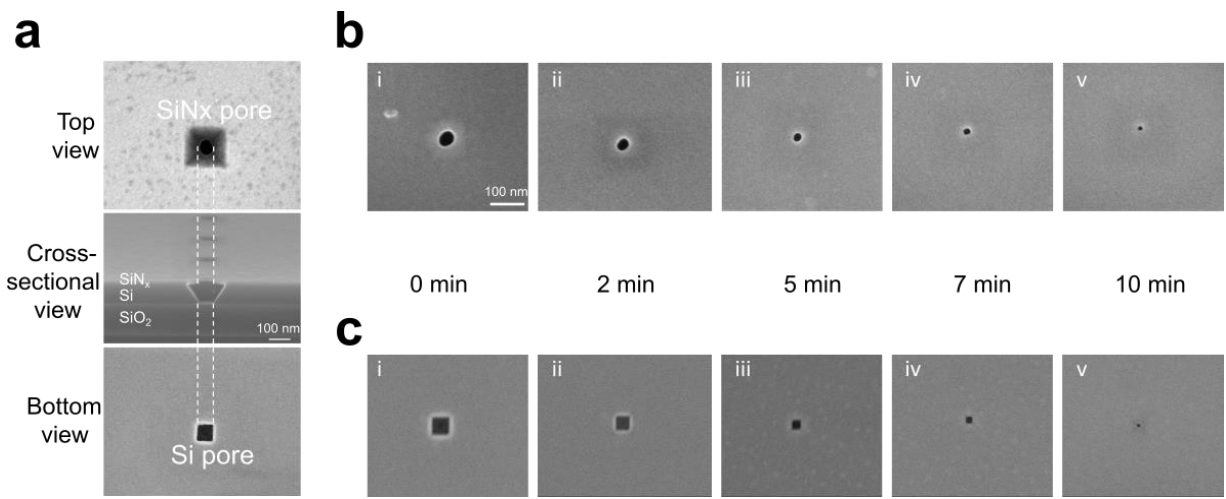

**Supplementary Fig. 3. Pore size reduction with real-time monitoring by SEM.** During SEM imaging, hydrocarbon was evaporated by an electron beam from a conductive carbon tab underneath the sample, allowing carbon deposition onto the nanopore surface to reduce pore size (a) SEM images of a nanocavity fabricated in an 88 nm thick Si membrane, showing top, cross-sectional, and bottom views. (b) Continuous reduction of a 43 nm SiNx pore under SEM at an accelerating voltage of 15 kV and magnification of 400 K: (i) 0 min, (ii) 2 min, (iii) 5 min, (iv) 7 min, (v) 10 min. (c) Continuous reduction of a 46 nm Si pore under the same conditions.

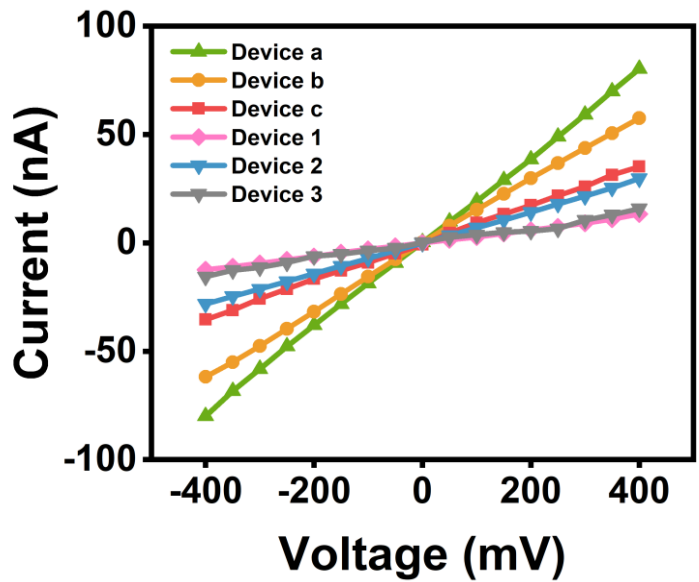

**Supplementary Fig. 4. Current-voltage (I-V) characteristics of the six devices in imaging buffer.**

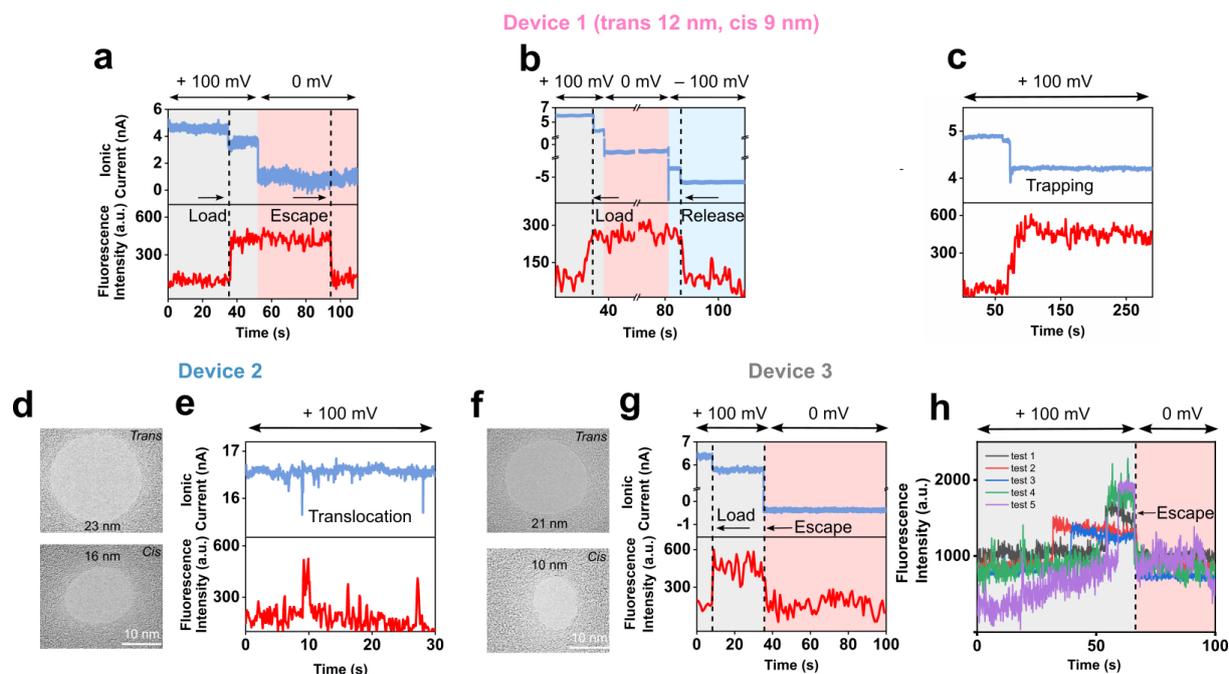

**Supplementary Fig. 5. Devices for trapping nucleosomes.** Ionic current and fluorescence intensity time traces showing detection of 1 nM fluorophore-labelled nucleosomes in imaging buffer. Data recorded by Device 1 (for TEM images, see Fig. 1b): (a) single nucleosome loading into the nanocavity at +100 mV (grey shading), trapping at 0 mV (pink shading), and subsequent spontaneous escape. (b) single nucleosome loading at +100 mV (grey shading), trapping at 0 mV (pink shading), and subsequent release at –100 mV (blue shading). (c) single nucleosome trapping at a constant +100 mV bias. Data recorded by Device 2: (d) TEM images showing 23 nm *trans* and 16 nm *cis* nanopores. (e) Nucleosome translocation events at a constant +100 mV bias. Data recorded by Device 3: (f) TEM images showing 21 nm *trans* and 10 nm *cis* nanopores. (g) Single nucleosome loading at +100 mV (grey shading) with immediate escape after removal of voltage (pink shading). (h) Independent replicate experiments showing instant fluorescence intensity decrease upon removal of voltage.

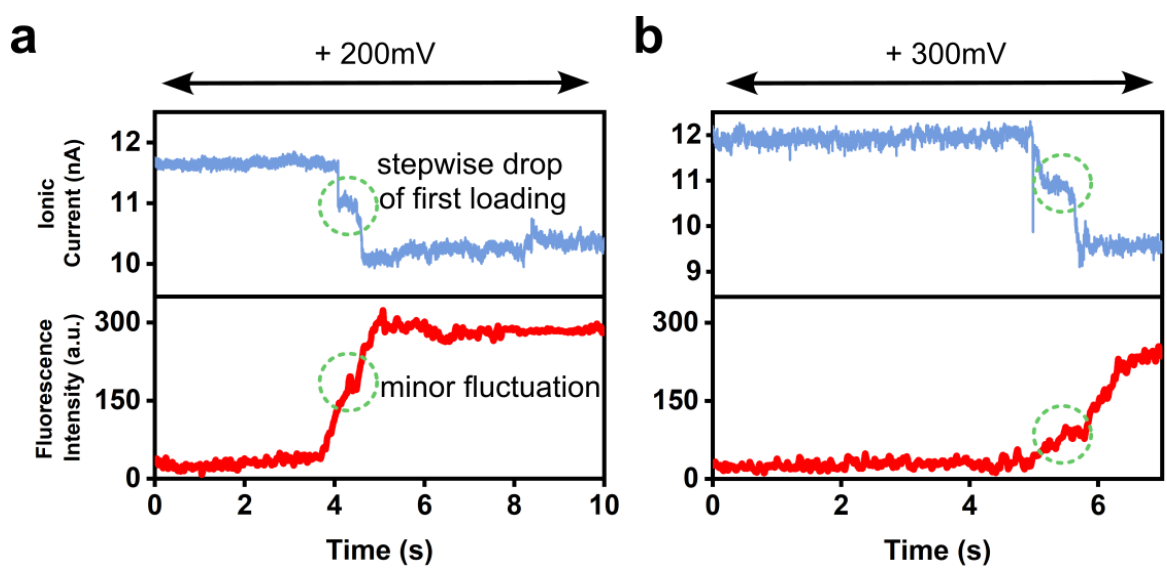

**Supplementary Fig. 6. Ionic current and fluorescence intensity time traces showing the sequential trapping of two fluorophore-labeled nucleosomes.** The applied voltage was held constantly at (a) +200 mV and (b) +300 mV.

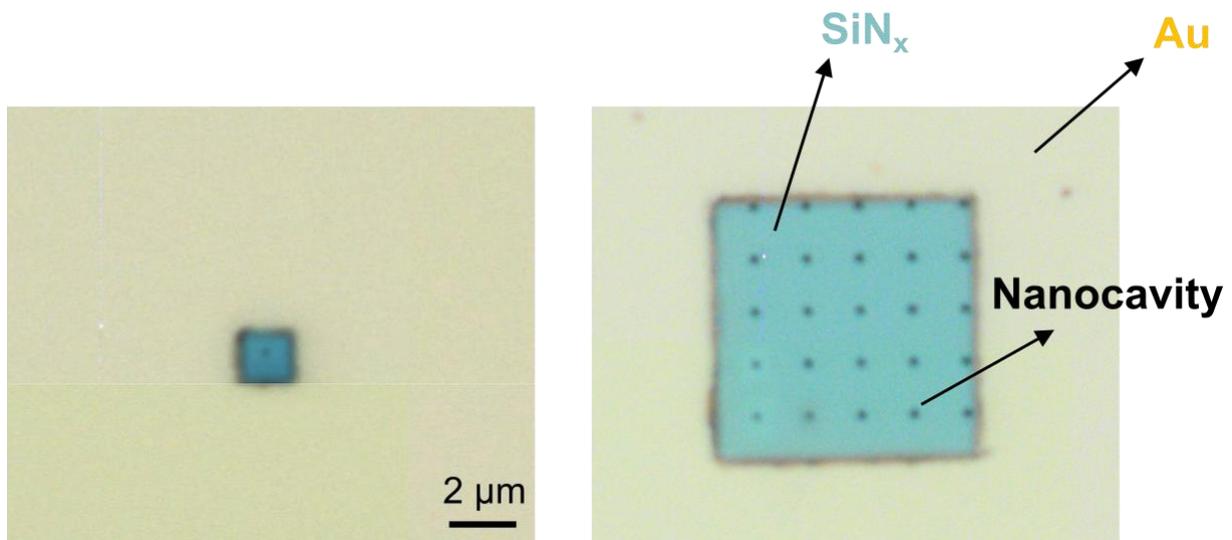

**Supplementary Fig. 7. 5 × 5 nanocavity array**. Optical micrographs from the *cis* side: single nanocavity (left) and 5x5 nanocavity array (right). Both devices are coated with a gold film.

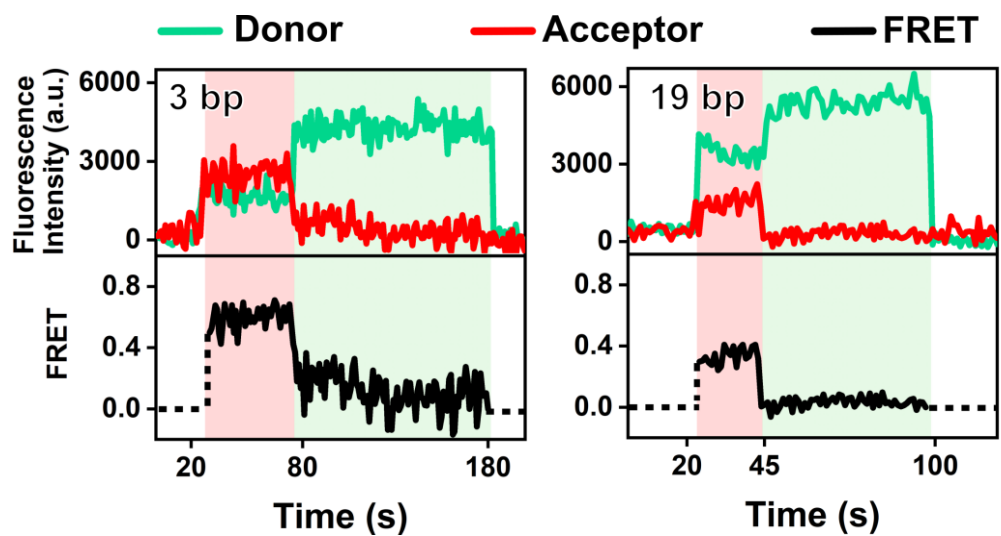

**Supplementary Fig. 8. Single-step photobleaching measurements.** Representative donor (green), acceptor (red) and FRET (black) time traces detected without the oxygen-scavenging system of 3-bp and 19-bp linker nucleosomes. The applied voltage was held constantly at +100 mV. Shaded areas: green indicates lower-FRET and red indicates higher-FRET states.

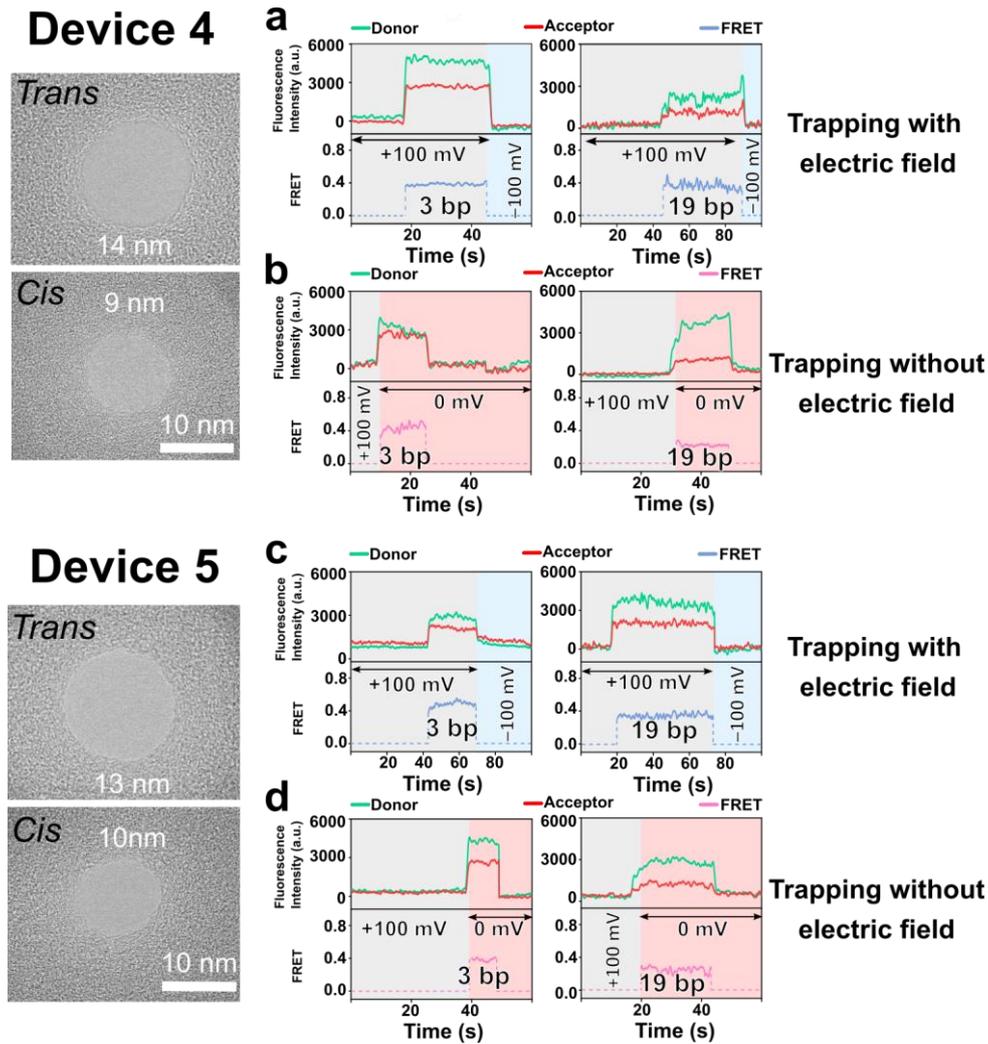

**Supplementary Fig. 9. Effect of the electric field on nucleosome conformation in other devices.** The left panel displays TEM images of Device 4 (14 nm *trans* and 9 nm *cis* nanopores) and Device 5 (13 nm *trans* and 10 nm *cis* nanopores). Representative time traces of donor (green), acceptor (red), and FRET (blue) signals for 3-bp and 19-bp nucleosomes recorded at +100 mV by (a) Device 4 and (c) Device 5. Representative time traces of donor (green), acceptor (red), and FRET (pink) signals for 3-bp and 19-bp nucleosomes recorded at 0 mV by (b) Device 4 and (d) Device 5. Shaded area color codes: grey for +100 mV, pink for 0 mV, and blue for −100 mV.

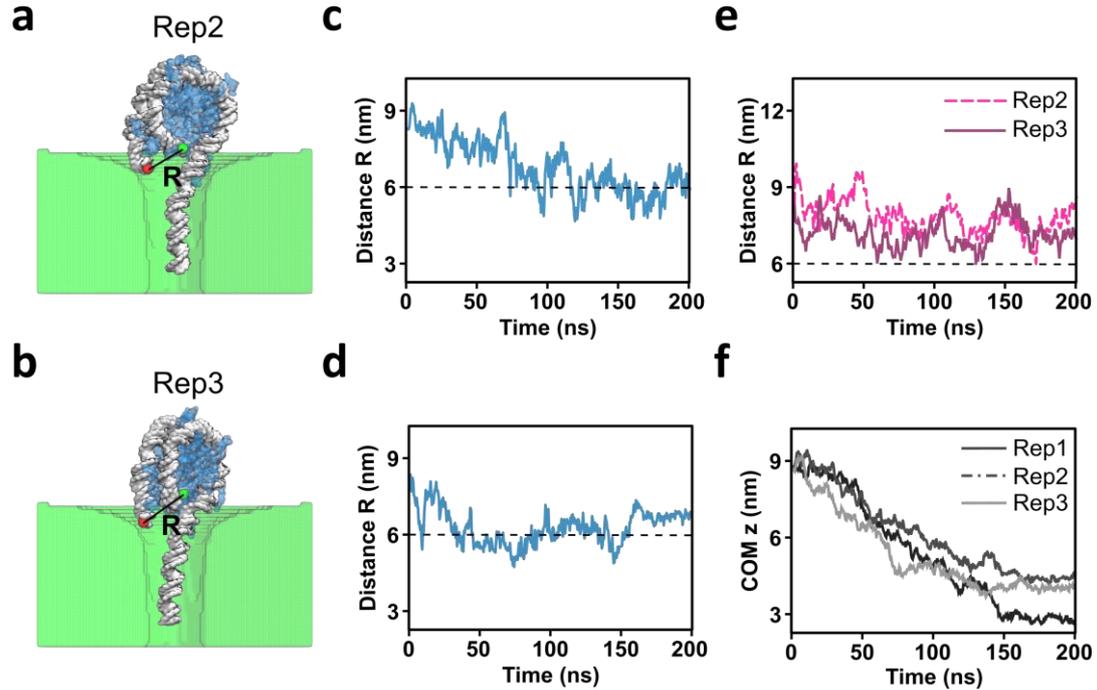

**Supplementary Fig. 10. Molecular dynamics simulation.** (a-b) Final frame of two independent MD simulations (Rep2 and Rep3) that are the replica of the system reported in Fig. 3E of the main manuscript. The MD are performed under an applied electric field E = (0, 0, Ez) corresponding to an electric potential difference of +100 mV across the nanopore. (c-d) Time evolution of the distance R between the fluorophore attachment sites for the replica Rep2 and Rep3, under a simulated voltage of +100 mV. The black dashed line marks the Förster distance ($R_0$), indicating the distance at which energy transfer is 50%. (e) Same distance R, without applied voltage or confinement, for other two independent simulations. (d) The center-of-mass (COM) z position of the histone core relative to the nanopore opening over time, for the systems under the applied electric field.

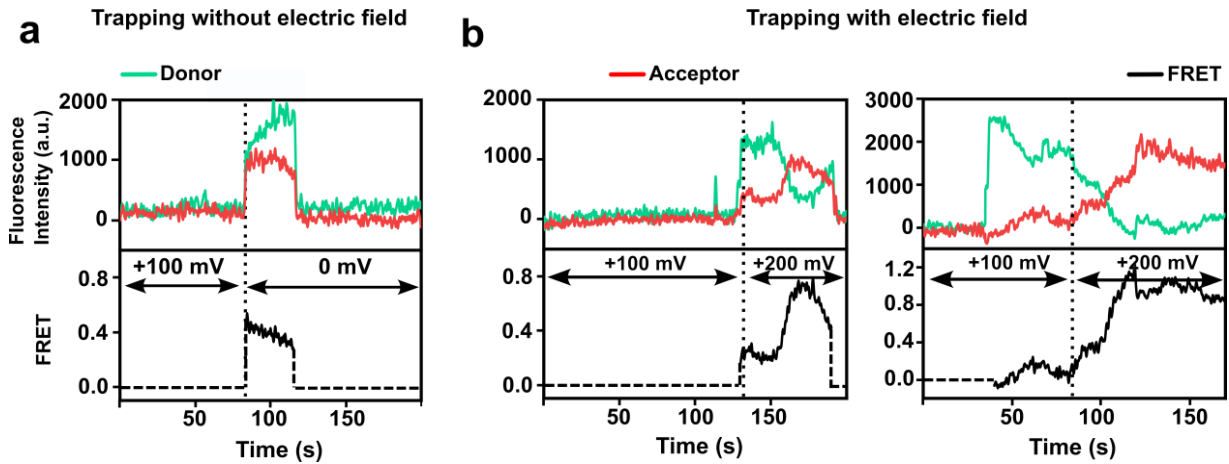

**Supplementary Fig. 11. Weak interaction between two nucleosomes labelled with Cy3 and Cy5, respectively.** Nucleosomes were loaded from pre-mixed solution. Representative time traces of donor fluorescence (green), acceptor fluorescence (red), and FRET efficiency (black) recorded (a) without electric field and (b) with electric field.

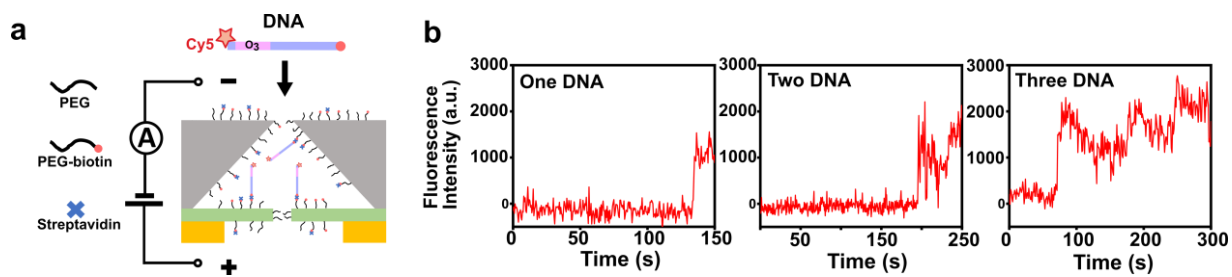

**Supplementary Fig. 12. DNA modified nanocavity.** (a) Schematic of the streptavidin-biotin-PEG surface immobilization and loading of Cy5-labelled DNA into nanocavity under +100 mV. (b) Representative time traces of Cy5 acceptor fluorescence (red) excited by a 638 nm laser, indicating various numbers of DNA molecules were loaded and bond with streptavidin inside the nanocavity.

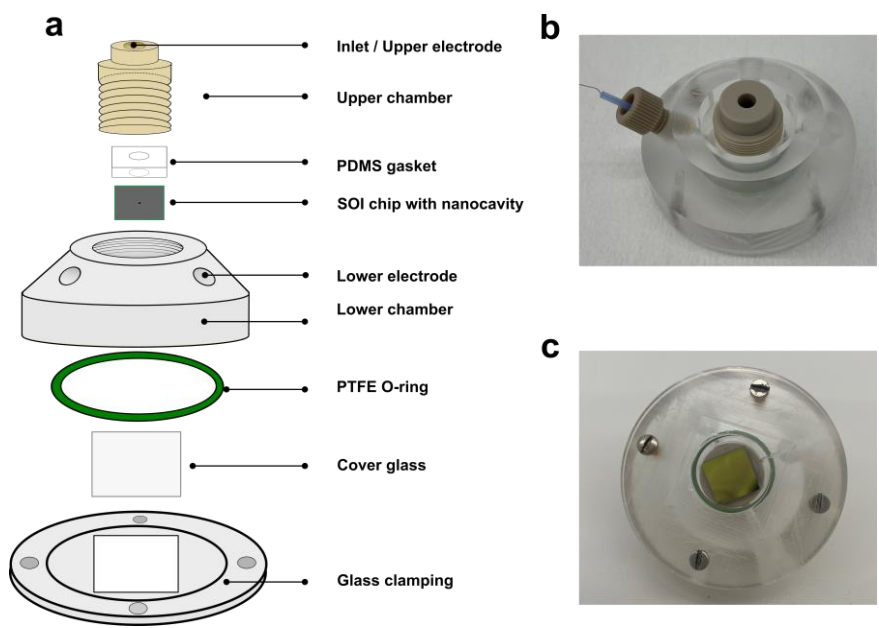

**Supplementary Fig. 13.** (a) Schematic of the custom-fabricated flow cell. (b) Top-view and (c) bottom-view optical images of the device.

**Supplementary Table 1. Sequences of LacI and O$_3$ construct for single-molecule experiments.**

| Name | Sequence |
|---|---|
| LacI-Cy3 | MKPVTLYDVAEYAGVSYQTVSRVVNQA**C(Cy3)**HVSAKTREKVEAAMAELNYIPNRVAQQLAGKQSLLIGVATSSLALHAPSQIVAAIKSRADQLGASVVVSMVERSGVEAAKAAVHNLLAQRVSGLIINYPLDDQDAIAVEAAATNVPALFLDVSDQTPINSIIFSHEDGTRLGVEHLVALGHQQIALLAGPLSSVSARLRLAGWHKYLTRNQIQPIAEREGDWSAMSGFQQTMQMLNEGIVPTAMLVANDQMALGAMRAITESGLRVGADISVVGYDDTEDSSCYIPPLTTIKQDFRLLGQTSVDRLLQLSQGQAVKGNQLLPVSLVKRKTTLAPNTQTHHHHHH |
| O$_3$ construct | Top strand:<br>5'-/5BioTinTEG/TCGTACTTCAAGTTTTGGGCGTGTCAAGTCCAAGGATTGCTCTGTATACTTAAAAACGACGTGGCAGTAAAGGGAACGCAAGACTCTCAATCGCGGCAGTGAGCGCAACGCAATTCCGAAAGCCT-3'<br>Bottom strand:<br>5'-AGGCT/iCy5/TCGGAATTGCGTTGCGCTCACTGCCGCGAATGAGAGTCTTGCGTTCCCTTTACTGCCACGTCGTTTTTAAGTATACAGAGCAATCCTTGGACTTGACACGCCCAAAACTTGAAGTACGA-3' |

**Supplementary Methods.**

**Molecular dynamics (MD) simulation.**

All the MD runs were carried out using GROMACS 2024[3] with a time step Δt = 2.0 fs. The force field used is the same as that employed by Winogradoff & Aksimentiev[4], based on Amber99sb-ILDN-PHI with bsc0 variant for DNA[5]. TIP3P model was used for water[6], and non-bonded corrections were applied for NaCl[7] and charged groups (CUFIX)[8]. A cutoff of 10 Å was used for the short-range nonbonded interactions. Particle mesh Ewald[9] method with a 1.6 Å spaced grid is used for long-range electrostatic interactions. A stochastic v-rescale thermostat[10] with a coupling constant of 0.1 ps applied to the entire system was used for all the simulations. Constraints were applied to bonded hydrogens using the SETTLE[11] algorithm for water and LINCS[12] for the other molecules. The initial velocities were generated from a Maxwell-Boltzmann distribution at 300 K. Periodic boundary conditions were applied in all three spatial dimensions. The membrane atoms were fixed in all directions and kept frozen in all the simulations. Production runs were performed at constant volume (NVT ensemble).

**Membrane preparation.** The membrane is made of uncharged hydrophilic dummy Lennard-Jones atoms (σ=0.37418 nm, ϵ=0.84 kJ/mol), having a simple cubic structure with atomic distance of 0.21 nm. A pore with a minor diameter of 6 nm is drilled through the membrane, using a smoothed function fitted from experimentally derived shape.

**Nucleosome preparation.** The complete structure of the histones, composing the nucleosome core protein, are taken from PDB 1KX5[13]. The dsDNA includes a 147-bp structure, wrapping the protein, elongated on the two sides with a shorter 19-bp and a longer 39-bp dsDNAs. The 147-bp structure is based on the Widom 601 sequence[14], and its structure is taken from PDB 3LZ0[15]. The 19-bp and 39-bp dsDNAs are generated and merged with ChimeraX[16]. The center of mass of the protein core of the nucleosome is placed at an initial distance of 9 nm from the membrane upper surface.

**Solvation and equilibration.** The final system is solvated into a rectangular box of 22x22x34 nm$^3$ and the total charge is neutralized by ionizing the system at 0.15M with NaCl, using GROMACS *solvate* and *genion* tools. The solvated system is then minimized for 1000 steps *via* descent gradient and then equilibrated to the correct temperature with an NPT simulation until the system reached a steady state volume (~5 ns). The nucleosome atoms were initially restrained (1000 kJ/mol/nm$^2$) and progressively halving the constraints every 500 ps during the first 2 ns; then the complex was completely free. Pressure coupling was conducted using a Parrinello-Rahman barostat[17] in a semi-isotropic manner, with separate coupling for the x/y plane and the z-axis. The reference pressure was set to 1 bar with a compressibility of $4.5\times10^{-5}$ bar$^{-1}$, and a coupling constant of 5.0 ps.

# References

1. Zeng, S., Wen, C., Solomon, P., Zhang, S.-L. & Zhang, Z. Rectification of protein translocation in truncated pyramidal nanopores. *Nat. Nanotechnol.* **14**, 1056–1062 (2019).

2. Zeng, S., Chinappi, M., Cecconi, F., Odijk, T. & Zhang, Z. DNA compaction and dynamic observation in a nanopore gated sub-attoliter silicon nanocavity. *Nanoscale* **14**, 12038–12047 (2022).

3. Abraham, M. J. *et al.* GROMACS: High performance molecular simulations through multi-level parallelism from laptops to supercomputers. *SoftwareX* **1–2**, 19–25 (2015).

4. Winogradoff, D. & Aksimentiev, A. Molecular Mechanism of Spontaneous Nucleosome Unraveling. *Journal of Molecular Biology* **431**, 323–335 (2019).

5. Pérez, A. *et al.* Refinement of the AMBER Force Field for Nucleic Acids: Improving the Description of α/γ Conformers. *Biophysical Journal* **92**, 3817–3829 (2007).

6. Jorgensen, W. L., Chandrasekhar, J., Madura, J. D., Impey, R. W. & Klein, M. L. Comparison of simple potential functions for simulating liquid water. *The Journal of Chemical Physics* **79**, 926–935 (1983).

7. Joung, I. S. & Cheatham III, T. E. Determination of alkali and halide monovalent ion parameters for use in explicitly solvated biomolecular simulations. *The journal of physical chemistry B* **112**, 9020–9041 (2008).

8. Yoo, J. & Aksimentiev, A. New tricks for old dogs: improving the accuracy of biomolecular force fields by pair-specific corrections to non-bonded interactions. *Phys. Chem. Chem. Phys.* **20**, 8432–8449 (2018).

9. Darden, T., York, D. & Pedersen, L. Particle mesh Ewald: An $N \cdot \log(N)$ method for Ewald sums in large systems. *The Journal of Chemical Physics* **98**, 10089–10092 (1993).

10. Bussi, G., Donadio, D. & Parrinello, M. Canonical sampling through velocity rescaling. *The Journal of Chemical Physics* **126**, 014101 (2007).

11. Miyamoto, S. & Kollman, P. A. Settle: An analytical version of the SHAKE and RATTLE algorithm for rigid water models. *Journal of computational chemistry* **13**, 952–962 (1992).

12. Hess, B., Bekker, H., Berendsen, H. J. & Fraaije, J. G. LINCS: a linear constraint solver for molecular simulations. *Journal of computational chemistry* **18**, 1463–1472 (1997).

13. Davey, C. A., Sargent, D. F., Luger, K., Maeder, A. W. & Richmond, T. J. Solvent mediated interactions in the structure of the nucleosome core particle at 1.9 Å resolution. *Journal of molecular biology* **319**, 1097–1113 (2002).

14. Lowary, P. & Widom, J. New DNA sequence rules for high affinity binding to histone octamer and sequence-directed nucleosome positioning. *Journal of molecular biology* **276**, 19–42 (1998).

15. Vasudevan, D., Chua, E. Y. & Davey, C. A. Crystal structures of nucleosome core particles containing the '601'strong positioning sequence. *Journal of molecular biology* **403**, 1–10 (2010).

16. Pettersen, E. F. *et al.* UCSF ChimeraX: Structure visualization for researchers, educators, and developers. *Protein science* **30**, 70–82 (2021).

17. Parrinello, M. & Rahman, A. Polymorphic transitions in single crystals: A new molecular dynamics method. *Journal of Applied physics* **52**, 7182–7190 (1981).